\documentclass[letterpaper]{JHEP3}

\usepackage{epsfig}

\def\endignore{}
\def\ignore #1\endignore{} 

\def\Box{{\hbox{$\sqcup$}\llap{\hbox{$\sqcap$}}}}

\def\be{\begin{equation}}
\def\ee{\end{equation}}
\def\bea{\begin{eqnarray}}
\def\eea{\end{eqnarray}}
\def\nn{\nonumber}

\def\exd{{\rm d}}
\def\pref#1{(\ref{#1})}
\def\endignore{}
\def\ignore #1\endignore{} 

\def\bd{\begin{displaymath}}
\def\ed{\end{diplaymath}}

\def\ie{{\it i.e.} }

\def\d{\mathrm{d}}

\def\cW{{\cal W}}
\def\cA{{\cal A}}

\def\cL{{\cal L}}

\def\cX{{\cal X}}
\def\cx{{\cal X}}
\def\cY{{\cal Y}}
\def\cy{{\cal Y}}
\def\cz{{\cal Z}}
\def\cZ{{\cal Z}}
\def\cU{{\cal U}}
\def\cV{{\cal V}}

\def\ssM{{\scriptscriptstyle M}}
\def\ssN{{\scriptscriptstyle N}}
\def\ssP{{\scriptscriptstyle P}}
\def\ssQ{{\scriptscriptstyle Q}}

\def\ba{\begin{eqnarray}}
\def\ea{\end{eqnarray}}
\def\be{\begin{equation}}
\def\ee{\end{equation}}

\def\mn{_{\mu \nu}}
\def\({\left(}
\def\){\right)}

\def\k{\kappa^2}

\newcommand{\eqref}[1]{(\ref{#1})}

\def\exd{{\rm d}}

\title{Kicking the Rugby Ball:\\Perturbations of 6D Gauged Chiral Supergravity}

\author{C.P.~Burgess,${}^{1,2}$ C.~de Rham,${}^{1,2,3}$ D.~Hoover,${}^3$
D.~Mason${}^1$ and A.J.~Tolley${}^{2,4}$\\
${}^1$ Dept. of Physics \& Astronomy, McMaster University,
Hamilton ON, L8S 4M1, Canada. \\
${}^2$ Perimeter Institute for Theoretical Physics, Waterloo ON, Canada. \\
${}^3$ Physics Department, McGill University, Montr\'eal QC, H3A
2T8, Canada. \\
${}^4$ Joseph Henry Laboratories, Princeton University, Princeton NJ, 08544, USA.\\
 \\
}

\date{}

\abstract {We analyze the axially-symmetric scalar perturbations
of 6D chiral gauged supergravity compactified on the general
warped geometries in the presence of two source branes.
We find all of the conical geometries are marginally stable
for normalizable perturbations
(in disagreement with some recent calculations) and the
nonconical for regular perturbations, even though none of them are
supersymmetric (apart from the
trivial Salam-Sezgin solution, for which there are no source
branes). The marginal direction is the one whose presence is
required by the classical scaling property of the field equations,
and all other modes have positive squared mass. In the special
case of the conical solutions, including (but not restricted to)
the unwarped `rugby-ball' solutions, we find closed-form
expressions for the mode functions in terms of Legendre and
Hypergeometric functions. In so doing we show how to match the
asymptotic near-brane form for the solution to the physics of the
source branes, and thereby how to physically interpret
perturbations which can be singular at the brane positions.}

\begin{document}

\section{Introduction}

Six dimensional supergravity continues to provide an instructive
laboratory for exploring different aspects of the cosmological
constant problem \cite{SLED1,SLED2}. It does so by building on the
earlier observation
\cite{5DSelfTune,5DSelfTunex,6DNonSUSYSelfTune,6DNonSUSYSelfTunex}
that higher-dimensional brane constructions can break the
implication that says that a nonzero 4D energy distribution must
necessarily generate a large 4D curvature. Once recast into a
higher-dimensional context the usual cosmological constant
problems \cite{CCproblem} become questions as to what are the
choices required for obtaining acceptably flat 4D space, and
identifying how natural these choices may be. Part of the appeal
of the 6D proposal is that it is very predictive, having
observational implications both for accelerator physics and for
tests of gravity in the lab and in astrophysics \cite{SLEDpheno},
in addition to its potential implications for late-time cosmology
\cite{ABRS2}.

In extra-dimensional models there are two separate kinds of
naturalness issues which must be addressed. The first type of
naturalness demands stability under renormalization for any
choices which are required to ensure the flatness of 4D space.
That is, any choice of brane positions or couplings which are required
for 4D flatness must be stable against integrating
out the ultraviolet part of the theory. Although
extra-dimensional models appear to make progress in understanding
this kind of stability \cite{UVSensitivity}, it remains open
whether they can completely resolve this part of the problem.

The second type of naturalness issue arises in extra-dimensional
models because these models generically relate the Dark Energy
density to the dynamics of the various moduli fields which
describe the geometry of the internal dimensions. Although it is
possible to obtain phenomenologically successful cosmology from
this kind of dynamics \cite{ABRS2}, the second naturalness issue
asks whether this requires unusually delicate choices for the
brane and bulk initial conditions. In particular one worries that
generic initial conditions --- or small perturbations to
specially-chosen ones --- might give rise to catastrophically fast
time evolution, leading to late-time cosmology which is unlike
what we see around us.

The present paper addresses
the second of these naturalness issues. Any understanding of this
issue requires knowledge of the time-dependent solutions of the
system. And since the time scales of interest are not slow
compared to those associated with the Kaluza Klein states, this
time evolution is intrinsically higher-dimensional and cannot be
accurately described within an effective 4D description
\cite{SLEDx}. Instead, what is required is an intrinsically
higher-dimensional description of the system's dynamics and its
response to various probes.

In a companion paper \cite{Scaling} we present a broad class of
time-dependent scaling solutions for 4D compactifications of 6D
supergravity. These solutions aid in understanding 
the evolution produced by generic brane properties by showing what the
late-time evolution of the system is likely to be for a wide class of
initial conditions. The present paper is aimed at a complementary part
of the problem of time-dependence. Instead of seeking the late-time
behaviour towards which the system evolves, we here study how small
perturbations can initiate the beginnings of time-dependent
behaviour. (See ref.~\cite{Waves} for a description of some of the
intermediate and transient time-dependent phenomena which lies between
these two cases.) We do so by investigating the linear perturbations
of a broad class of static solutions to 6D supergravity, in order to
see how they respond to arbitrary small perturbations.
In this analysis, we use a Kaluza Klein decomposition which includes
both the modes oscillating in time as well as the zero-mass non-oscillating ones.

Our main result is that these systems are marginally stable for
normalizable perturbations around conical geometries, and for regular
perturbations around nonconical ones.
(We also examine a class of physically
motivated 
non-regular perturbations, for which our stability
analysis is inconclusive.) The stability is only marginal because
of the generic scaling property which the 6D supergravity
solutions have. (It is the absence of an energy barrier in this
marginal direction which allows the development of the scaling
solutions studied in ref.~\cite{Scaling}.) We identify the set of
equations which govern the time evolution of the system after all
of the constraints have been used to eliminate some of the field
fluctuations, and reproduce the form found by earlier workers
\cite{LP}. But we differ in the conclusions which we draw
regarding stability from these equations, for reasons we explain
in detail in the Appendix.

We begin our discussion in \S\ref{sec 2} with a reminder of the
field equations and some of the known static background
configurations. This is followed in \S\ref{sec 3} by the
derivation of the equations governing linearized perturbations in
comoving gauge, while the same derivation in longitudinal gauge is
performed in appendix~\ref{sec long gauge} for comparison with
previous results. \S\ref{sec 4} then solves analytically the
linearized equations in the conical case and identifies the
asymptotic behaviour in the more general non-conical case. We then
give in \S\ref{sec stability}  several general arguments in favour
of the stability of the modes examined and relate them with the
boundary conditions in \S\ref{sec 4}. 
After a brief
summary of conclusions -- \S\ref{sec conclusion} -- four
appendices contain: details of the analysis in longitudinal gauge,
with the full stability argument; a detailed comparison of why our
stability conclusions differ from some previous results; the
relation between some of the results in various gauges; and a
summary of some properties of the special functions used in the
explicit solutions.

\section{Field Equations and Background Solutions}
\label{sec 2}
We begin in this section by summarizing the relevant field
equations, as well as the static background solutions about which
we shall perturb.

\subsection{Field equations}

The bosonic part of the Lagrangian density for 6D chiral gauged
supergravity is given (for the case of vanishing hyperscalars ---
$\Phi^a = 0$) by \cite{NS}\footnote{The curvature conventions used
here are those of Weinberg's book \cite{GandC}, for which all
curvature tensors differ by an overall sign relative to those of
MTW \cite{MTW}.}
\bea \label{6DSugraAction}
    \frac{{\cal L}}{\sqrt{-g}} &=& - \frac{1}{2 \kappa^2} \,
    g^{\ssM\ssN} \Bigl[ R_{\ssM\ssN} + \partial_\ssM \varphi \,
    \partial_\ssN \varphi \Bigr] - \frac{2g^2}{\kappa^4} \; e^\varphi \nn\\
    && \qquad\qquad - \frac14 \, e^{-\varphi} \, F_{\ssM\ssN} F^{\ssM\ssN}
    - \frac{1}{2 \cdot 3!} \, e^{-2\varphi} \, G_{\ssM\ssN\ssP}
    G^{\ssM\ssN\ssP} \,,
\eea
where $F_{\ssM\ssN} = \partial_\ssM A_\ssN - \partial_\ssN A_\ssM$
and $G_{\ssM\ssN\ssP} = \partial_\ssM B_{\ssN\ssP} + \partial_\ssN
B_{\ssP\ssM} + \partial_\ssP B_{\ssM\ssN} + (A_\ssP F_{\ssM\ssN}
\, \hbox{terms})$. The coupling constants $g$ and $\kappa$
respectively have dimension (mass)${}^{-1}$ and (mass)${}^{-2}$.

The field equations obtained from this action are:
\bea \label{6DSugraEqns}
    \Box \, \varphi + \frac{\kappa^2}{6} \, e^{-2 \varphi} \;
    G_{\ssM\ssN\ssP} G^{\ssM\ssN\ssP} +
    \frac{\kappa^2}{4} \, e^{-\varphi} \; F_{\ssM\ssN} F^{\ssM\ssN}
    - \frac{2 \,g^{2}}{\kappa^2} \, e^{\varphi} &=& 0
    \qquad \hbox{(dilaton)}\nn \\
    D_\ssM \Bigl(e^{ -2 \varphi} \, G^{\ssM\ssN\ssP} \Bigr) &=& 0 \qquad
    \hbox{(2-Form)} \nn\\
    D_\ssM \Bigl(e^{ - \varphi} \, F^{\ssM\ssN} \Bigr) + e^{-2 \varphi}
    \, G^{\ssM\ssN\ssP} F_{\ssM\ssP} &=& 0 \qquad \hbox{(Maxwell)}\nn\\
    R_{\ssM\ssN} + \partial_\ssM\varphi \, \partial_\ssN\varphi
    + \frac{\kappa^2}{2} \, e^{-2 \varphi}
    \; G_{\ssM\ssP\ssQ} {G_\ssN}^{\ssP\ssQ} + \kappa^2 e^{- \varphi} \;
    F_{\ssM\ssP} {F_\ssN}^\ssP
    + \frac12 \, (\Box\,
    \varphi)\, g_{\ssM\ssN} &=& 0 \,.\quad\;\; \hbox{(Einstein)}\nn\\
\eea

An important feature of these equations is their classical scaling
property. This property states that given any solution to these
equations another can be obtained by making the replacement
\be
    e^\varphi \to \lambda e^\varphi \qquad \hbox{and} \qquad
    g_{\ssM\ssN} \to \lambda^{-1} g_{\ssM\ssN} \,,
\ee
with all other fields held fixed. This property follows from the
fact that the action, eq.~\pref{6DSugraAction}, scales under this
transformation according to $S \to \lambda^{-2} S$ \cite{witten}.

The remainder of the paper deals with solutions to these equations
for which two dimensions are compact and four are not. To this end
denote the six coordinates $x^\ssM$, $M = 0,1,2,3,4,5$, as $x^\ssM
= \{t,x,y,z,\eta,\theta \}$. We denote 4D coordinates by $x^\mu$,
$\mu = 0,1,2,3$, where $x^\mu = \{t,x,y,z\}$, and 2D coordinates
of the extra dimensions by $x^m$, $m = 4,5$, where $x^m = \{
\eta,\theta \}$. When required, the three spatial coordinates of
the observable four dimensions are denoted $x^i, i = 1,2,3$ and so
$x^i = \{x,y,z \}$.

\subsection{Ans\"atze}

In later sections our interest is in metrics of the form
\be \label{diagmetricform}
    \exd s^2 =  e^{2 a} \,
    \eta_{\mu\nu} \, \exd x^\mu \exd x^\nu
    + e^{2 v} \, \exd \eta^2 + e^{2 b}
    \exd \theta^2
\ee
and consider these component functions, as well as $\varphi$ and
$A_\theta$, to depend only on the coordinates $\eta$ and $t$. (In
what follows we sometimes generalize this assumption to allow
dependence on $\eta$ and $x^\mu$.) Denoting differentiation with
respect to $\eta$ and $t$ by primes and dots, the field equations
for these functions then reduce to the following set of coupled
partial differential equations.

\medskip\noindent The Maxwell equation is:
\be
    -e^{2(v-a)} \left[ \ddot A_\theta + (2 \dot a + \dot v - \dot b
    - \dot \varphi) \dot A_\theta \right] + A_\theta''
    + ( 4a' - v' -b' -\varphi') A_\theta'
    = 0 \,.
\ee
The dilaton equation is:
\bea
    &&- e^{2(v-a)} \left[\ddot \varphi + (2 \dot a +
    \dot v + \dot b) \dot\varphi \right]
    + \varphi'' + (4a' -v' +b' ) \varphi'\nn\\
    && \qquad\qquad\qquad - \frac12
    \, e^{ -2a + 2v - 2b -\varphi} (\dot A_\theta)^2 +
    \frac12 \, e^{- 2b-\varphi} (A_\theta')^2
    - \frac{2 g^2}{\kappa^2} \, e^{2v+\varphi} =0\,,
\eea
The $(t\eta)$ Einstein equation is
\be
    3\, \dot a' + \dot b' -3 \, \dot v \, a' + \dot b
    \, b' - \dot b \, a' - \dot v \, b' +
    \dot\varphi \varphi' + e^{-2b-\varphi}
    \dot A_\theta A_\theta' = 0 \,.
\ee
The $(tt)$ Einstein equation is:
\bea
    && - e^{2(v-a)} \Bigl[ 3\, \ddot a  + \ddot b + \ddot v
    + (\dot v)^2 + (\dot b)^2 + (\dot \varphi)^2
    - \dot a ( \dot v +  \dot b)\Bigr]
    + a'' + a'(4 \, a' - v' + b') \nn \\
    &&\qquad\qquad\qquad - \frac{3\kappa^2}{4}
    \, e^{-2a + 2v - 2b -\varphi} (\dot A_\theta)^2
    - \frac{\kappa^2}{4} \,
    e^{- 2b-\varphi} (A_\theta')^2
    + \frac{g^2}{\kappa^2}\, e^{2v+ \varphi}
    = 0 \,.
\eea
The $(\eta\eta)$ Einstein equation is:
\bea
    &&-e^{2(v - a)} \Bigl[\ddot v + \dot v (
    2 \, \dot a + \dot v + \dot b) \Bigr] + 4\, a''
    + b'' + 4 \, (a')^2 + (b')^2 - v' (4 \, a' + b')
    + (\varphi')^2  \nn\\
    &&\qquad\qquad\qquad
    + \frac{\kappa^2}{4}
    \, e^{-2a + 2v -2b -\varphi} (\dot A_\theta)^2
    + \frac{3\kappa^2}{4} \,
    e^{- 2b-\varphi} (A_\theta')^2
    + \frac{g^2}{\kappa^2} e^{2v+ \varphi} = 0 \,.
\eea
The $(\theta\theta)$ Einstein equation is:
\bea
    &&- e^{2(v - a)} \Bigl[\ddot b + \dot b (2\, \dot a
    + \dot v + \dot b) \Bigr]
    +  b'' + b' (4\, a'
    -v' + b' )  \nn\\
    && \qquad\qquad\qquad - \frac{3\kappa^2}{4}
    \, e^{-2a + 2v - 2b  -\varphi} (\dot A_\theta)^2
    + \frac{3\kappa^2}{4} \,
    e^{- 2b-\varphi} (A_\theta')^2
    + \frac{g^2}{\kappa^2}\, e^{2v+ \varphi} = 0 \,.
\eea
The $(ij)$ Einstein equation is:
\bea
    &&-e^{2(v - a)} \,
    \Bigl[\ddot a + \dot a ( 2\, \dot a
    + \dot v + \dot b) \Bigr]  + a''
    + a' (4\, a' - v' + b') \nn\\
    &&\qquad\qquad\qquad + \frac{\kappa^2}{4}
    \, e^{- 2a + 2v - 2b-\varphi} (\dot A_\theta)^2 -
    \frac{\kappa^2}{4} \, e^{- 2b-\varphi } (A_\theta')^2
    + \frac{g^2}{\kappa^2} \, e^{2v+\varphi}  =0 \,.
\eea

\subsection{Maximally symmetric 4D compactifications}

Many explicit compactifications of the above field equations to
four dimensions have been constructed over the years, starting 20
years ago with the Salam-Sezgin spherical solution \cite{NS}.
These now include compactifications to flat 4D space on unwarped,
rugby-ball solutions \cite{SLED1}, as well as warped
axially-symmetric internal dimensions having conical \cite{SLED2},
and more general \cite{GGP,GGPplus,Parameswaran:2006db} singularities at the positions
of two source branes. More recent generalizations have found
warped de Sitter 4D geometries \cite{6DdSSUSY}, as well as
configurations for which the hyperscalars and 3-form fluxes are
nontrivial \cite{HypersNonzero}.

Since we wish to perturb about the 4D flat solutions, we briefly
summarize their properties here.\footnote{The conventions of
ref.~\cite{GGP} may be obtained from ours by taking $R_{MN} \to -
R_{MN}$, $\varphi \to -\varphi/2$ and $\kappa^2 = 1/2$, while
those of \cite{SLED2} differ from those here only by the choice
$\kappa^2 = 1$.} These solutions may be written
\be
    e^a = \cW \,, \qquad
    e^v = \cA \cW^4 \qquad \hbox{and} \qquad
    e^b = \cA \,,
\ee
and so the bulk fields become
\bea \label{GGPFields}
    \exd s^2 &=& \cW^2(\eta) \, \eta_{\mu\nu} \, \exd x^\mu \exd x^\nu
    + \cA^2 (\eta) \Bigl[ \cW^8(\eta) \, \exd \eta^2 +
    \exd \theta^2 \Bigr] \nn\\
    F_{\eta\theta} &=& \left( \frac{q \cA^2 }{\cW^2} \right)
    e^{-\lambda_3 \eta} \qquad \hbox{and} \qquad
    e^{-\varphi} = \cW^2 \, e^{\lambda_3 \eta} \,,
\eea
where
\bea \label{GGPSolutions}
    \cW^4 &=& \left( \frac{\kappa^2 q \lambda_2}{2 g \lambda_1}
    \right) \frac{\cosh[\lambda_1(\eta -
    \eta_1)]}{\cosh[\lambda_2(\eta-\eta_2)]} \nn\\
    \cA^{-4} &=& \left( \frac{2 \kappa^2 q^3 g}{\lambda_1^3
    \lambda_2} \right) e^{-2\lambda_3 \eta} \cosh^3[
    \lambda_1(\eta -\eta_1)] \cosh[ \lambda_2(\eta - \eta_2)] \,,
\eea
where the field equations imply the constraint $\lambda_2^2 =
\lambda_1^2 + \lambda_3^2$.

There are two particularly interesting special cases of these
solutions. The first is obtained by taking $\lambda_3=0$, in which
case the resulting geometries have purely conical singularities
\cite{GGPplus}. The second comes by taking both $\lambda_3 = 0$
and $\eta_1=\eta_2$, in which case the conical geometry is also
unwarped, since $\varphi = \varphi_0$ and $\cW = \cW_0$ are
constants. In this latter case changing variables to proper
distance, $\exd \rho = \cA \cW^4 \exd \eta$, the Maxwell field
becomes
\be
    F_{\rho\theta} = \partial_\rho A_\theta
    = \pm \frac{2g}{\kappa^2} \, e^{\varphi_0} \, B(\rho) \,,
\ee
where $B(\rho) \equiv e^{b(\rho)}$ satisfies
\be
    \frac{B''}{B} = b'' + (b')^2
    = - \frac{4g^2}{\kappa^2} \, e^{\varphi_0} \,,
\ee
with solution $B(\rho) = B_0 \sin(\rho/\rho_0)$ with $\rho_0 =
\frac12 \kappa e^{-\varphi_0/2}/g$. These are the unwarped,
rugby-ball generalizations \cite{SLED1} of the original spherical
Salam-Sezgin solution \cite{NS} to include the back-reaction of
two branes, located at the sphere's north and south poles.

In what follows we also use the extremely useful change of
variables used in ref.~\cite{6DdSSUSY},
\ba
    \varphi=\frac12 \(\cx-\cy-2\cz\),
    \hspace{10pt} \log \cW=\frac 1 4 \(\cy-\cx\)
    \hspace{5pt} \rm{and} \hspace{5pt} \log \cA=\frac 1 4
    \(3\cx+\cy+2\cz\),
\ea
and so
\ba
    e^{-\cal{X}}&=&e^{-\varphi/2}\cA^{-1}=\frac{\kappa q}{\lambda_1}\cosh
    \left[\lambda_1
    (\eta-\eta_1)\right]\\
    e^{-\cal{Y}}&=&e^{-\varphi/2}\cA^{-1}\cW^{-4}=\frac{2 g}{\kappa \lambda_2}
    \cosh \left[\lambda_2 (\eta-\eta_2)\right]\\
    e^{-\cal{Z}}&=&e^{\varphi}\cW^2=e^{-\lambda_3 \eta} \,.
\ea
It is straightforward to check that these are related by the
Hamiltonian constraint (for evolution in the $\eta$ direction):
\ba
    - \frac{4 g^2}{ \k} \, e^{2\cy}
    +\k q^2 e^{2 \cx}+\cx'{}^2-\cy'{}^2+\cz'{}^2=0 \,.
    \label{constraint}
\ea

\subsection{Asymptotic forms}

The solutions to these equations describe geometries which
typically become singular within the bulk. Because of the assumed
axial symmetry this can occur at two (or fewer) points, which we
can choose to place at $\eta = \pm \infty$. These are interpreted
as being the positions of the 3-branes which source the
corresponding field configuration. We require an explicit form for
the asymptotic behaviour near these singularities in order to
determine the boundary conditions in our later linearized
perturbation analysis. We therefore pause here to outline what
this asymptotic behaviour is.

If we use proper distance as the radial coordinate near the branes
we have $\exd s^2 = \hat g_{ab} \, \exd x^a \exd x^b + \exd
\rho^2$, and we imagine the brane position to be given by $\rho =
0$.
For maximally-symmetric 4D geometries the asymptotic form of the
bulk fields in the $\rho \to 0$ limit is then generically given by
a power law form \cite{6DdSSUSY}
\bea \label{asymptoticfields}
    &&\exd s^2 \sim [c_a (H\rho)^{\alpha}]^2\, q_{\mu\nu} \,
    \exd x^\mu \exd x^\nu
    +  \exd \rho^2 + [c_\theta (H\rho)^{\beta}]^2 \exd \theta^2
    \nn\\
     &&\qquad\qquad e^{\varphi} \sim c_\phi (H\rho)^p
        \quad \hbox{and} \quad
        F^{\rho\theta} \sim c_f \, (H\rho)^\gamma\,,
\eea
where $\alpha$, $\beta$, $p$, $\gamma$, $c_a$, $c_\theta$,
$c_\phi$ and $c_f$ are constants, $H$ is an arbitrary dimensionful
parameter, and $q_{\mu\nu}$ denotes the metric of 4D flat, de
Sitter or anti-de Sitter space. For time-independent fields the
field equations impose the following two Kasner-like constraints
amongst the powers $\alpha$, $\beta$, $\gamma$ and $p$:
\be \label{PowerFieldEqns}
    4 \,\alpha^2 + \beta^2 + p^2 = 4\, \alpha +
    \beta = 1 \quad \hbox{and} \quad
    \gamma = p -1 \,.
\ee
Imposing these 3 conditions on the 4 powers leaves 1 undetermined,
which we can choose to be $p$. Together with the prefactor,
$c_\theta$, this power can be related to the two physical choices
which are available for the source branes: their tension and
dilaton coupling \cite{Scaling,GGPplus,NavSant,Peloso:2006cq}. For example,
explicit calculation using the solutions given above gives
\be \label{GGPPowers}
    \alpha_\pm = \frac{\lambda_2 - \lambda_1}{5 \lambda_2 - \lambda_1
    \mp 2 \lambda_3} \,,  \quad
    \beta_\pm = \frac{\lambda_2 + 3 \lambda_1 \mp 2\lambda_3}{5
    \lambda_2 - \lambda_1 \mp 2 \lambda_3} \quad
    \hbox{and} \quad
    p_\pm = -\frac{2( \lambda_2 - \lambda_1 \mp 2 \lambda_3)}{5
    \lambda_2 - \lambda_1
    \mp 2 \lambda_3}  \,,
\ee
for the asymptotic form as $\eta \to \pm \infty$. As is easily
verified, these satisfy eqs.~\pref{PowerFieldEqns} given above.

For time-dependent geometries it is useful to generalize the above
to allow different asymptotics for the spatial and temporal
components of the metric, replacing eq.~\pref{asymptoticfields}
with:
\be \label{asymptoticfields2}
    \exd s^2 \sim -[c_w (H\rho)^{\omega}]^2\,
    \exd t^2 +  [c_a (H\rho)^{\alpha}]^2\,\, g_{ij}
    \exd x^i \exd x^j
    +  \exd \rho^2 + [c_\theta (H\rho)^{\beta}]^2 \exd \theta^2
    \,,
\ee
with the powers now related by
\be \label{PowerFieldEqns2}
    \omega^2 + 3 \,\alpha^2 + \beta^2 + p^2 = \omega + 3\, \alpha +
    \beta = 1 \quad \hbox{and} \quad
    \gamma = p -1 \,.
\ee
The additional parameter appearing in this asymptotic form is
related to the freedom to separately specify the energy density
and pressure on the source branes.

\section{Linearization}
\label{sec 3}
In this section we set up the equations which govern
axially-symmetric perturbations about the solutions described
above which transform as scalars in 4D. The restriction to axially
symmetric perturbations does not limit the ensuing stability
analysis because the most unstable modes are generally the most
symmetric, since any angular dependence contributes positively to
the corresponding Kaluza-Klein mass.

\subsection{Symmetries and gauge choices}

The most general 4D scalar perturbations have the form
\bea
    \exd s^2 &=& e^{2a} \Bigl[ \eta_{\mu\nu} + M_{,\, \mu\nu} \Bigr]
    \exd x^\mu \exd x^\nu + N_{m,\mu} \, \exd x^\mu \exd x^m
    + g_{mn} \,\exd x^m \exd x^n \nn\\
    B &=& B_{\mu\nu} \, \exd x^\mu \wedge \exd x^\nu
    + B_{m,\mu} \, \exd x^\mu \wedge \exd x^m
    + B_{mn} \, \exd x^m \wedge \exd x^n \nn\\
    \hbox{and} \qquad A &=& \Omega_{,\mu} \, \exd x^\mu
    + A_{m} \, \exd x^m  \,,
\eea
where in four dimensions $B_{\mu\nu}$ dualizes to a 4D scalar
$\zeta$ through a relation of the form $H_{\mu\nu\lambda} \propto
\epsilon_{\mu\nu\lambda\rho} \partial^\rho \zeta$. We are free to
use gauge symmetries to locally set $N_m = B_m = \Omega = 0$.

\subsubsection*{Discrete symmetry and mode mixing}

As has been pointed out by earlier analyses \cite{LP,Jim}, for the
purposes of a stability analysis it is not necessary to keep all
of the remaining perturbations. To see why, notice that since our
interest is in fluctuations depending only on $t$, $x^i$and $\eta$, the
absence of a dependence on the coordinate $\theta$
implies also a symmetry under the reflection, $\theta \to - \theta$.
If $A_\theta$ is nonzero in the background
configuration (as is the case when a flux $F_{\eta\theta}$ is
turned on in the extra dimensions), this is only a symmetry if we
also independently require $A_\ssM \to - A_\ssM$.

This symmetry ensures that the linearized fluctuations can be
divided into two classes, according to whether or not they are
even or odd under these reflections. In particular we have
\bea
    \{\delta a,\delta g_{\theta\theta},\delta g_{\eta\eta},
    \delta \varphi, \delta A_\theta \} \qquad &&(\hbox{even}) \nn\\
    \{\delta g_{\eta\theta}, \delta \zeta, \delta B_{\eta\theta},
    \delta A_\eta \} \qquad &&(\hbox{odd})
    \,.
\eea
Since the symmetry guarantees that it is consistent with all of
the equations of motion to set the odd fluctuations to vanish,
these two categories of fluctuations cannot mix with one another
at the linearized level.
We take advantage of this fact to choose $\delta g_{\eta\theta} =
\delta \zeta = \delta B_{\eta\theta} = \delta A_\eta = 0$. Again,
the omission of these modes does not restrict the stability
analysis which follows.

We are led in this way to the following ansatz for the metric and
other bulk fields
\bea
    \exd s^2 &=& e^{2a} \Bigl( \eta_{\mu\nu} + M_{,\, \mu\nu} \Bigr)
    \exd x^\mu \exd x^\nu + e^{2v} \, \exd \eta^2 + e^{2b} \, \exd
    \theta^2 \nn\\
    A &=& A_\theta \, \exd \theta \qquad \hbox{and} \qquad B = 0 \,.
\eea
To linearize the field equations about a static background we
write
\bea
    &&a(\eta,x) = a_0(\eta) + A(\eta,x) \,, \quad
    v(\eta,x) = v_0(\eta) + V(\eta,x) \,, \quad
    b(\eta,x) = b_0(\eta) + B(\eta,x) \,, \nn\\
    &&\qquad\qquad
    \varphi(\eta,x) = \varphi_0(\eta) + \Phi(\eta,x) \quad
    \hbox{and} \quad
    A_\theta(\eta,x) = a_\theta(\eta) + \cA_\theta(\eta,x) \,,
\eea
where we allow the fluctuations to depend on all 4 coordinates,
$x^\mu$, and the background field configurations are given as
above:
\bea
    &&e^{a_0} = \cW(\eta) \,, \quad
    e^{v_0} = \cA(\eta) \cW^4(\eta) \,, \quad
    e^{b_0} = \cA(\eta) \,, \nn\\
    &&\qquad
    e^{-\varphi_0} = \cW^2(\eta) \, e^{\lambda_3 \eta}
    \quad \hbox{and} \quad
    a_\theta' = \frac{q \cA^2(\eta)}{\cW^2(\eta)} \, e^{-\lambda_3 \eta} \,.
\eea
For some of the discussion to follow it is useful to choose proper
distance computed using the background metric as a coordinate within
the extra dimensions, as we did for the rugby ball solution above.
We reserve the variable $\rho$ for proper distance, and it is
related explicitly to $\eta$ by $\exd \rho = e^{v_0(\eta)} \exd
\eta = \cA \cW^4 \, \exd \eta$.
Notice that this choice of coordinates can be made independently
of the gauge choice we make on the linearized fluctuations, which
we now describe.

\subsubsection*{Three useful gauge choices}

Finally, it is also not necessary to keep all of $A$, $M$, $V$ and
$B$ independent, because one combination of these can be set to
zero using an appropriate coordinate choice. There are three gauge
choices which we use in what follows.

\begin{itemize}
\item {\it Comoving} ($c$) gauge: defined by the condition
$\cA_\theta^{(c)} =0$, leading to
\be \label{com gauge}
    \d s^2 = e^{2a_0} e^{2A^{(c)}} \, \eta\mn \d x^\mu \d x^\nu
    + e^{2v_0} e^{2V^{(c)}} \, \d \eta^2
    + e^{2b_0} e^{2B^{(c)}} \, \d \theta^2
    + 2N_{, \mu} \, \d \eta \d x^\mu \,,
\ee
together with $A = a_\theta \d \theta$ and $\varphi = \varphi_0 +
\Phi^{(c)}$. Notice that this gauge does not fall completely into
the ansatz of eq.~\pref{diagmetricform}, because the coordinate
transformation required to get into this form does not preserve
the condition $\delta A_\theta = 0$. This is the origin of the new
metric variable, $\delta g_{\eta \mu}$, in the above expression.
\item {\it Gaussian Normal} ($GN$) gauge, defined by $V^{(GN)} =
0$ and so:
\ba
    \d s^2 = e^{2 a_0} \(e^{2A^{(GN)}} \eta\mn+M^{(GN)}_{,\mu\nu}\)
    \d x^\mu \d x^\nu + e^{2b_0} e^{2B^{(GN)}} \d \theta^2
    + \d \rho^2 \,,
\ea
together with $\varphi=\varphi_0+\Phi^{(GN)}$ and
$A_\theta=a_\theta+\cA_\theta^{(GN)}$. It is obviously convenient
also to use background-metric proper distance, $\rho$ defined via $d\rho=e^{v_0}d\eta$.%
\item {\it Longitudinal} ($l$) gauge: defined by $M=0$ and so
\ba
\label{long gauge 2}
    \d s^2=e^{2 a_0} e^{2A^{(l)}}\eta\mn \d x^\mu \d x^\nu
    + e^{2b_0}e^{2B^{(l)}} \d \theta^2 + e^{2V^{(l)}}
    \d \rho^2,
\ea
together with $\varphi = \varphi_0 + \Phi^{(l)}$ and $A_\theta =
a_\theta + \cA_\theta^{(l)}$. Notice that we again choose to use
background-metric proper distance, $\rho$, as the radial
coordinate.
\end{itemize}

Which of these gauges is most convenient depends on the goal of
the calculation. Comoving gauge is physically intuitive (since one
follows perturbations along hypersurfaces of constant $A_\theta$),
and has the enormous benefit that the field equations completely
decouple in this gauge, making the linearization analysis much
easier. Longitudinal gauge is convenient for making contact with
earlier calculations in the literature, since this is the gauge
which has often been used. Finally, GN gauge is useful for
analyzing the boundary conditions which fluctuations must satisfy
near the source branes.

It is clearly useful to be able to transform between expressions
obtained in these three gauges, so we next briefly summarize the
main formulae which are required. To change from longitudinal to
comoving gauge requires the transformation
$x_{(l)}^{\scriptscriptstyle M} = x_{(c)}^{\scriptscriptstyle M} +
\xi^{\scriptscriptstyle M}$, with $\xi^{\scriptscriptstyle M} =
\varepsilon \, \delta_\rho^{\scriptscriptstyle M}$ so that
$\rho_{(l)} = \rho_{(c)} + \varepsilon$, with $\varepsilon =
\cA^{(l)}_\theta / \partial_\rho a_\theta$ required to ensure
$\cA^{(c)}_\theta = 0$. This implies that the various fields in
longitudinal gauge are related to the ones in comoving gauge by:
\ba
    &A^{(l)} = A^{(c)} + \varepsilon \, \partial_\rho a_0 \,,
    \hspace{10pt}
    B^{(l)} = B^{(c)} + \varepsilon \, \partial_\rho b_0 \,,
    \hspace{10pt}
    V^{(l)} = V^{(c)} + \partial_\rho \varepsilon \,,
    \hspace{10pt}\label{long to com} \\
    &N^{(c)} = - e^{4a_0+b_0} \varepsilon \,,
    \hspace{10pt}
    \Phi^{(l)} = \Phi^{(c)} + \varepsilon \partial_\rho
    \varphi_0 \,. \nn
\ea

Similarly, to go from comoving to GN gauge, we perform the
transformations $\rho_{(c)} = \rho_{(GN)} + \varepsilon$ with
$\partial_\rho \varepsilon = V^{(c)}$ chosen to ensure $V^{(GN)} =
0$. The further transformation, $x^\mu_{(c)} = x^\mu_{(GN)} +
\partial^\mu \epsilon$, with $\partial_\rho \epsilon=
 e^{-(6a_0+b_0)} N^{(c)} - e^{-2a_0} \varepsilon$ is also required
 in order to ensure the vanishing of $N^{(GN)}$. Quantities
in comoving gauge are then related to those in GN gauge by
\ba \label{com to GN}
    & A^{(c)}= A^{(GN)}+ \varepsilon \partial_\rho a_0 \,,
    \hspace{10pt}
    M^{(GN)} = -2 \epsilon,
    \hspace{10pt}
    B^{(c)} = B^{(GN)} + \varepsilon \partial_\rho b_0  \nn\\
    & \Phi^{(c)}=\Phi^{(GN)} +\varepsilon \partial_\rho
    \varphi_0, \hspace{10pt} \cA_\theta^{(GN)} = -\varepsilon
    \partial_\rho a_\theta.
\ea

\subsection{Linearized equations in comoving gauge}

We now work out the perturbed equations in comoving gauge, since
in this gauge we are able to decouple the fluctuations even when
perturbing about non-conical background solutions.
We use as the coordinate $\eta$ in this section, with the
derivative $\exd/\exd\eta$ denoted by primes.

We adopt the following notation for the perturbations
$A^{(c)} = -\Psi/2$ and $V^{(c)} = \xi/2$, and so
\be
    \exd s^2 = e^{(\cY - \cX)/2} e^{-\Psi} \, \eta_{\mu\nu} \,
    \exd x^\mu \exd x^\nu + e^{(3\cX + \cY + 2\cZ)/2} \Bigl[
    e^{2(\cY - \cX)} e^{\xi} \, \exd \eta^2 + e^{2B} \, \exd
    \theta^2 \Bigr]  + 2 N_{,\mu} \exd \eta \, \exd x^{\mu} \,,
\ee
while $e^{\varphi} = e^{(\cX - \cY - 2\cZ)/2} e^{\Phi}$ and
$F_{\eta\theta} = a_\theta' = q e^{-\lambda_3\eta} e^{2\cX +
\cZ}$.
In terms of these variables, and Fourier transforming with respect
to time and space\footnote{
We emphasize that $\omega$ is the
mass of the corresponding Kaluza-Klein mode and should be understood
as the eigenvalue of the operator $\Box$. In this analysis, we do
not make the assumption that the modes have a specific time-oscillatory behavior
of the form $e^{i k_0 t}$.
Thus this analysis includes both time-oscillating modes, and
non-oscillating ones which correspond to zero-mass Kaluza Klein modes.}
so $\Box = \eta^{\mu\nu}\partial_\mu
\partial_\nu= \omega^2$, the Maxwell equation is
\ba \label{EqMaxwell}
    2 \,\omega^2 e^{(\cx-\cy)/2} N + 2B' + 2\Phi' + \xi' + 4\Psi' = 0
    \,;
\ea
the dilaton equation is
\ba
    \Phi'' + \omega^2 e^{2\cy+\cz} \Phi - \frac12 \(\k q^2 e^{2\cx}
    +\frac{4g^2}{\k} e^{2\cy} \) \Phi &=& \varphi_0'
    \(\frac12 \xi' + 2 \Psi' - B' + \omega^2 e^{(\cx-\cy)/2} N\) \nn \\
    && \qquad\qquad + 
    \frac{2g^2}{\k} \, e^{2\cy} \xi + \k q^2 e^{2\cx} B \,;
\ea
the $(\mu \eta)$ Einstein equation is
\ba \label{Eqzeta}
    2 (2 \cx' + \cz') B -(2\cy'+\cz') \xi + 4\varphi_0' \Phi + 4 B'-6\Psi'=0\,,
\ea
while the other off-diagonal term in the Einstein equation (the
combination $(ii)-(jj)$, with $i\ne j$) is
\ba
    e^{\frac 12(\cx-5\cy-2\cz)} \left[ 2N'+(\cx'-\cy') N \right]
    -2B-\xi+2\Psi=0.
\ea
A combination of the $(\eta\eta)$ Einstein equation, the dilaton
equation of motion and the trace of the Einstein equation which
involves only first derivatives with respect to $\eta$ is
\ba \label{EqN}
    \frac{\omega^2}{2}\, e^{(\cx-\cy)/2}(2\cy'+\cz') N
    -\frac{2g^2}{\k} \, e^{2\cy} \xi
    + \frac{3\omega^2}{2}\, e^{2\cy+\cz} \Psi
    +(2 \cy'+\cz') \Psi' \nn \hspace{100pt}\\
    -\frac12 \(\k q^2 e^{2\cx} + \frac{4g^2}{\k} \, e^{2\cy} \) \Phi
    + \varphi_0' \Phi' - \k q^2  e^{2\cx} B - \omega^2 e^{2\cy+\cz} B
    +(\cx'-\cy') B' = 0 \,. \hspace{20pt}
\ea
The $(\theta \theta)$ Einstein equation is similarly
\ba
    (2B''+\Phi'') + \omega^2 e^{2\cy+\cz} (2B+\Phi)
    -2 \k q^2 e^{2\cx} (2B+\Phi) + 2 \cX' B' \hspace{60pt} \nn\\
    = \cx' \(4\Psi'+\xi' + 2\, \omega^2 e^{(\cx-\cy)/2} N\),
\ea
and finally, the $(\mu \mu)$ Einstein equation is
\ba
    \(\Phi'' + \omega^2 e^{2\cy+\cz} \Phi\)
    - \(\Psi'' + \omega^2 e^{2\cy+\cz} \Psi - 2\cz' \Psi'\)
    =\cz' \( B'- \frac12\, \xi' - \omega^2 e^{(\cx-\cy)/2} N \) .
\ea

We have here a set of seven equations for five variables ($\xi$,
$N$, $\Psi$, $B$ and $\Phi$). Two of these equations appear as
constraints and can be used to fix the lapse and shift functions:
using the $(\mu \eta)$ equation \eqref{Eqzeta} and the $(\eta
\eta)$ equation \eqref{EqN}, we can express both $\xi$ and
$\omega^2 N$ in terms of the three remaining variables $\Psi, B$
and $\Phi$, leading to five remaining equations for three
variables. However these remaining equations are not independent
since the Bianchi identities ensure that two of
them
are redundant. One
can be expressed as a linear combination of the others, whilst the
derivative of the Maxwell equation \eqref{EqMaxwell} is simply the derivative of
\eqref{Eqzeta} and a combination of the other equations.

We arrive in this way with three equations for three unknowns,
which can be written as
\ba \label{Eq B}
    && \hspace{-30pt}B'' + \omega^2 e^{2\cy+\cz} B
    + \frac14 \(6\cx' + 2\cy' + 4\cz' + \frac{16\,g^2}{\k}
    \, \frac{e^{2\cy}}{2\cy'+\cz'} \) B' \nn \\
    &&+\frac18 \(-12 \k q^2 e^{2\cx} + \frac{16\,g^2}{\k} \,
    e^{2\cy} \frac{2\cx'+\cz'}{2\cy'+\cz'} \) B
    +\frac{1}{4}\(3\cx'+\cy'+2 \cz'\)\Phi' \\
    &&\hspace{30pt}+\frac{1}{16}\(-12 \k q^2 e^{2\cx}
    +\frac{16\,g^2}{\k} \, e^{2\cy} \frac{2\cx'-3\cz'}{2\cy'+\cz'} \)
    \Phi -\frac{6\,g^2}{\k} \,\frac{e^{2\cy}}{2\cy'+\cz'} \Psi'=0
    \,,\nn
\ea
\ba \label{Eq Psi}
    &&\hspace{-30pt}\Psi'' + \omega^2 e^{2\cy+\cz} \Psi
    +\frac{12\, g^2}{\k} \, \frac{e^{2\cy}}{\(2\cy'+\cz'\)} \Psi'
    +\frac12 (\cx'-\cy')\(\Phi'+2B'\)
    -\frac{8\,g^2}{\k}\frac{e^{2\cy}}{2\cy'+\cz'} B' \\
    &&\hspace{10pt}
    -\frac{\k q^2}{2} \, e^{2\cx} \(\Phi+2B\)
    -\frac{2\,g^2}{\k} \, \frac{e^{2\cy}}{2\cy'+\cz'}
    \left[(2\cx'-3\cz') \Phi + 2(2\cx'+\cz')B \right] = 0 \,, \nn
\ea
and
\ba \label{Eq Phi}
    &&\hspace{-30pt} \Phi'' + \omega^2 e^{2\cy+\cz} \Phi
    +\varphi_0' (\Phi'+2B')
    +\frac{4\,g^2}{\k} \, \frac{e^{2\cy}}{2\cy'+\cz'} \(3\Psi'-2B'\) \\
    &&\hspace{5pt}
    -\frac{\k q^2}{2}\, e^{2\cx} \(\Phi+2B\)
    -\frac{2\,g^2}{\k} \, \frac{e^{2\cy}}{2\cy'+\cz'}
    \left[(2\cx'-3\cz') \Phi + 2(2\cx'+\cz') B \right]
    =0 \,. \nn
\ea

These equations dramatically simplify once the change of variables
$\( \Phi ,B, \Psi\) \rightarrow \(\tilde{f}, \tilde{\chi}, \tilde
\psi\)$ is performed, with
\ba
%
    \Phi &=& \frac{2\, \cU}{\kappa q} \, e^{-\cx} \tilde{f} - 2B \nn \\
    B &=& \frac{\cV}{\sqrt3 \,\cU} \, \tilde{\chi}
    -\frac12\, \Psi + \frac{2 \,\cU^2
    + \cx' \cz'}{2 \,\kappa q \,\cU } \, e^{-\cx} \tilde{f}   \label{change variable 2}\\
    \Psi &=& \frac{2\cy'+\cz'}{4\cV} \tilde \psi
    + \frac{2\cU^2 - 3\cy'\cz'}{4\sqrt{3} \,\cU\cV} \tilde\chi
    + \frac{\cU^2 -\cx'\cy'}{2 \,\kappa q \,\cU } \, e^{-\cx} \tilde{f}
    \,, \nn
\ea
where use of the background field equations shows that the
nominally field-dependent quantities $\cU$ and $\cV$ are really
both constants:
\ba
    \cU^2 &\equiv& \cx'^2 + \k q^2 e^{2\cx} = \lambda_1^2 \nn\\
    \cV^2 &\equiv& \cx'^2 + \k q^2 e^{2\cx} + \frac34 \cz'^2
    = \lambda_1^2 + \frac34 \lambda_3^2 \,.
\ea
Notice that for conical backgrounds $\cU = \cV = \lambda_1$. In
terms of these new variables the three field equations ---
eqs.~\pref{Eq B}, \pref{Eq Psi} and \pref{Eq Phi} --- take the
remarkably simple, decoupled form
\ba
    &&\tilde{\chi }'' + \omega^2 e^{2\cy+\cz} \tilde{\chi } = 0
    \label{eq talpha}\\
    &&\tilde{f}'' + \omega^2 e^{2\cy+\cz} \tilde{f} - \cU^2 \tilde{f} = 0
    \label{eq tphi} \\
    &&\tilde{\psi}'' + \omega^2 e^{2\cy+\cz} \tilde{\psi}
    -\frac{32 \, g^2}{\k} \, \frac{\cV^2 e^{2\cy}}{\(2\cy'+\cz'\)^2}
     \tilde{\psi} = 0 \,.
    \label{eq tpsi}
%
%
\ea

Because these equations decouple from one another each can be
solved separately, and presents no problem for numerical analysis.
Exact solutions are also available in the case of backgrounds
having conical singularities (including but not restricted to the
rugby ball geometries), as shall be described in detail in later
sections. Given the existence of variables for which the equations
decouple in comoving gauge, one might wonder what the
corresponding variables are in other gauges. These variables can
be obtained by explicitly performing the required change of gauge,
but because the results are not very simple we relegate their
description to Appendix \ref{app decoupled variables}.

It is noteworthy for later purposes that the decoupled equations
(\ref{eq talpha} -- \ref{eq tpsi}) are independent of the
background variable $\cx$ (and hence also of the parameter
$\eta_1$). Consequently the linearized equations of motion for the
new variables $\tilde{\chi }, \tilde{f}$ and $\tilde \psi$ are
identical for a general conical background geometry (for which
$\cz'=0$) and a rugby ball geometry (for which $\cz'=0$ and
$\cx'=\cy'$). However once the solutions are inserted into the
expressions for the metric fluctuations using relations
\eqref{change variable 2}, the results do depend on $\cx$ and so
the full fluctuations `know' about the complete background
geometry.

\section{Properties of Solutions}
\label{sec 4}

Up to this point there has been no restriction on the form of the
background geometry. In this section we specialize to special
conical backgrounds for which it is possible to solve the above
linearized equations in closed form. We then return to the more
general non-conical geometry in order to discuss the asymptotic
behaviour of the solutions in the near-brane limit. The section
closes with an outline of how this asymptotic form is related to
the physical properties of the branes which source the bulk
geometries.

\subsection{Analytic Solutions for General conical backgrounds}
\label{exactsol}

We now concentrate on the behaviour of the perturbations to
solutions having only conical singularities, which are defined by
the condition $\lambda_3 = 0$ (and so $\cz'\equiv 0$). Performing
the change of variable $z = \tanh \left[ \lambda_2 \(\eta - \eta_2
\) \right]$, the perturbation equations become
\ba
    &&\frac{\d^2 \tilde{\chi
    }}{\d z^2}-\(\frac{2z}{1-z^2}\)\frac{\d
    \tilde{\chi }}{\d z}+\(\frac{\mu^2}{1-z^2}\)\tilde{\chi }=0\\
    && \frac{\d^2 \tilde{\psi}}{\d z^2}-\(\frac{2z}{1-z^2}\)\frac{\d
    \tilde{\psi}}{\d z}+\(\frac{1}{1-z^2}\)\(\mu^2-\frac{2
    }{z^2}\)\tilde{\psi}=0\\
    &&\frac{\d^2 \tilde{f}}{\d z^2}-\(\frac{2z}{1-z^2}\)\frac{\d
    \tilde{f}}{\d z}+\(\frac{1}{1-z^2}\)\(\mu^2-\frac{1
    }{1-z^2}\) \tilde{f}=0,
\ea
with $\mu^2=\omega^2 \k/4 g^2$, and $\omega$ is the 4D mode mass.
Notice that these equations are symmetric under $z\rightarrow -z$,
which corresponds to reflections of $\eta$ around $\eta = \eta_2$.
The reason for this symmetry is clear in the special case of the
rugby-ball solutions (for which $\eta_1 = \eta_2$), since it then
corresponds to reflections of the spherical geometry about its
equator. In the more general warped conical geometries this
symmetry instead follows from the above-mentioned circumstance
that the equations are completely independent of $\eta_1$, and so
take the same form for general conical geometries as they do for
the rugby ball.

Defining $\nu=\frac12 \left[ -1 + \sqrt{1+ 4\mu^2} \right]$ --- or
equivalently writing $\mu^2 = \nu(\nu+1)$ --- the solutions to
these equations become
\ba
    \tilde \chi &=& C_1 \, P_\nu(z)+ C_2 \,  \hbox{Re}
    \, Q_\nu(z)\\
    &=& C_1 F\left[-\nu,1+\nu;1;\frac 12(1-z)\right]
    + C_2 \sqrt{\pi}\,
    \frac{\Gamma(\nu+1)}{\Gamma(\nu+\frac 32)} \, \hbox{Re} \,\frac{1}{(2z)^{\nu+1}}\,
    F\left[\frac
    12(1+\nu),1+\frac \nu 2;\nu +\frac 32; z^{-2}\right]
    \nn\\
    \tilde \psi &=& C_3 \, z^{-1}
    F\left[-\frac 1 2 (1+\nu),\frac \nu 2 ;-\frac 1 2;z^2\right]
    +C_4 \, z^{2}
    F\left[1-\frac \nu 2 ,\frac 1 2 \(3+\nu\);\frac 5 2;z^2\right]\\
    \tilde f&=&\frac{1}{\sqrt{1-z^2}} \(
    C_5 \, F\left[\frac \nu 2,-\frac 12 (1+\nu);\frac 1 2; z^2 \right]
    + C_6\ z \ F\left[\frac 1 2(1+\nu),-\frac \nu 2 ;\frac 3 2; z^2 \right] \),
\ea
where the $C_i, \, i = 1,..,6$ are integration constants and
$F(a,b;c;z)$ is the hypergeometric function. Neglecting the
overall normalization of all three modes, we see that the above
expressions provide a 9-parameter family of solutions.
Notice also that these solutions can be decomposed into symmetric
and anti-symmetric combinations under the symmetry $z \rightarrow -z$, and
this is already explicit for the functions 
$\tilde\psi$ and $\tilde f$.

We see that the mode energies satisfy
\be \label{KKspectrum}
    \omega^2 = \frac{\nu(\nu + 1)}{r_0^2} \,,
\ee
with $1/r_0 = 2g/\kappa$, and so the spectrum depends on the
values which are allowed for the parameter $\nu$. Notice that for
real $\nu$ we have $\omega^2 \ge 0$ except for the interval $-1 <
\nu < 0$. As usual, the values which are allowed for $\nu$ are
related to the behaviour which is demanded of the solutions at the
boundaries, which in the present instance corresponds to the
near-brane limits $z \to \pm 1$. For this reason we next explore
the asymptotic form taken by the solutions in the near-brane
limit.

\subsection{Asymptotic Forms}

We next examine three aspects of the near-brane limit. In this
section we first explore the asymptotic forms taken by the
solutions just constructed for fluctuations about backgrounds
having only conical singularities. We then see how these
asymptotic forms generalize to the broader situation where the
background has more generic singularities. A preliminary
connection is drawn in the next section between this asymptotic
behaviour and the properties of the source branes.

\subsubsection*{Conical Backgrounds}

The asymptotic form of the solutions found above near the brane
positions at $z = \pm 1$ may be found by using the properties of
the Hypergeometric functions given in appendix \ref{app special
functions}. This shows that for generic $\nu$, $\tilde \chi$ and
$\tilde \psi$ diverge logarithmically as one approaches the
branes, whereas $\tilde f$ generically diverges as $(1-z^2)^{-1/2}
\sim\cosh[\lambda_2(\eta-\eta_2)] \sim\rho^{-1}$.

As usual, less singular asymptotic behaviour can be possible for
those $\nu$ for which the hypergeometric series terminates, which
for $F(a,b;c;z)$ occurs when either $a$ or $b$ vanishes.
Inspection of the explicit solutions then shows that termination
happens when $\nu$ is quantized to be a non-negative integer: $\nu
= \ell = 0,1,2,\dots$. Notice that the resulting quantization of
KK mass, from eq.~\pref{KKspectrum}, implies in this case
$\omega^2 \ge 0$.

For $\nu = \ell$ one -- but not both -- of the two Hypergeometric
functions appearing for each of $\tilde\chi$, $\tilde\psi$ and
$\tilde f$ terminates, allowing a less singular near-brane limit
to be obtained through an appropriate choice for the integration
constants $C_i$. For instance, if we require --- as we shall argue
on the grounds of normalizability below --- that all fluctuations
must be less singular than $(1-z^2)^{-1}$, then this is possible
for $\tilde f$  only if $\nu = \ell$ and one of $C_5$ or $C_6$
vanishes (which one depends on whether $\ell$ is even or
odd).\footnote{We give explicit expressions for some of the lowest
of these nonsingular modes for $\tilde\psi$ and $\tilde f$ in
Appendix A for longitudinal gauge.} Once this has been done the
functions $\tilde\chi$ and $\tilde\psi$ can still diverge
logarithmically as $z \to \pm 1$, leaving a 4-parameter family of
potentially logarithmically singular solutions (up to overall
normalization).

Completely regular solutions are also possible when $\nu = \ell$,
provided that one of $C_1$ or $C_2$ and one of $C_3$ or $C_4$ also
vanishes. For instance, $\tilde\chi$ is regular everywhere if $\nu
= \ell$ and $C_2 = 0$ since in this case the solution $P_\ell$
degenerates to the Legendre polynomials. This leaves a 2-parameter
family of nonsingular solutions (up to overall normalization).

\subsubsection*{Non-conical backgrounds}

Although it is difficult to find explicit solutions in the generic
case where the background has non-conical singularities. It is
possible to use the linearized equations to estimate the
asymptotic behaviour of the mode functions as $\eta\rightarrow \pm
\infty$ (or $z\rightarrow \pm 1$). Repeating the steps which led
to the equations of the previous section for this more general
case, we instead find
\ba \label{revasympt}
    &&\frac{\d^2 \tilde{\chi }}{\d
    z^2}-\(\frac{2z}{1-z^2}\)\frac{\d \tilde{\chi }}{\d
    z}+\frac{\tilde{\mu}^2}{(1-z^2)}\(\frac{1+z}{1-z}\)^{\delta \lambda}
    \tilde{\chi }=0\\
    && \frac{\d^2 \tilde{\psi}}{\d z^2}-\(\frac{2z}{1-z^2}\)\frac{\d
    \tilde{\psi}}{\d z}+\(\frac{1}{1-z^2}\)\left[\frac{\tilde
    \mu^2}{(1-z^2)}\(\frac{1+z}{1-z}\)^{\delta \lambda}
    -\frac{2 \(1-\delta\lambda^2\)
    }{(z-\delta\lambda)^2}\right]\tilde{\psi}=0\\
    &&\frac{\d^2 \tilde{f}}{\d z^2}-\(\frac{2z}{1-z^2}\)\frac{\d
    \tilde{f}}{\d z}+\(\frac{1}{1-z^2}\)\left[
    \frac{\tilde \mu^2}{(1-z^2)}\(\frac{1+z}{1-z}\)^{\delta \lambda}
    -\frac{1-4 \delta \lambda^2}{1-z^2}\right] \tilde{f}=0,
\ea
with $\tilde{\mu}^2=\mu^2\, e^{\lambda_3 \eta_2}$ and $\delta
\lambda=\lambda_3/2 \lambda_2$.

Since $|\delta \lambda|<1/2$ the term in eq.~\pref{revasympt}
which is proportional to $\tilde \chi$ remains subdominant
compared to the terms in $\(1-z^2\)^{-1} \exd\tilde \chi/\exd z$
and $\exd^2\tilde \chi/\exd z^2$, such that the asymptotic
behaviour of $\tilde \chi$ still remains logarithmic near the
brane. Similar arguments show that $\tilde \psi$ and $\tilde f$
also keep the same asymptotic behaviour as in the conical case.

\subsubsection*{Normalizability}

A restriction on the asymptotic forms follows from the condition
that the mode functions be normalizable, as we now explore. As we
show in detail in the next section, the equations of motion for
each of the decoupled modes can be derived from a reduced action,
eq.~\eqref{Action decoupled}. In this action all modes, $u$, have
the same kinetic term, $e^{2\cy+\cz} u \Box u$, where $e^{2\cy +
\cz} = \cA^2 \cW^6$, and so the definition of the inner product
can be taken to be the same for all three of $\tilde \chi, \tilde
\psi$ and $\tilde f$, having the form:
\ba
    \langle u,v \rangle=\frac i 2
    \int_\Sigma \d^3 x \, \d\theta\, \d \eta \, \cA^2\cW^6 \(u
    \partial_t v^*-u^* \partial_t v\)\nn,
\ea
for any two modes $u$ and $v$. Here $\Sigma$ denotes a surface of
constant $t$. The norm of a mode is therefore defined by
\ba
    \langle u,u \rangle = \omega \int_\Sigma \d^3 x \,
    \d\theta \,\d \eta \, \cA^2\cW^6  u u^*
    = \omega \int_\Sigma \d^3 x \,\d\theta\, \d \rho \, \cA \cW^2  u u^*,
\ea
where we use $\exd\rho/\exd\eta = \cA\cW^4$ and $\partial_t u =
-i\omega u$.

Our interest is in the conditions placed on the asymptotic
behaviour of linearized modes by the convergence of these
integrals. Given the asymptotic form, eq.~\pref{asymptoticfields},
of the background metric, with powers $\alpha_0$, $\beta_0$, $p_0$
{\it etc.}, we have $\cA \cW^2 \propto \rho^{\beta_0 + 2\alpha_0}
\propto \rho^{1-2\alpha_0}$ in the near-brane limit ($\rho \to
0$). In order to be normalizable, modes should therefore vary as
$u \propto \rho^\upsilon$ with power $\upsilon >  -1 + \alpha_0$
as $\rho \to 0$. In particular, for a conical background with
$\alpha_0 = 0$ and $\beta_0 = 1$, modes must diverge less rapidly
than $1/\rho$ near both branes.

\subsection{Boundary conditions and brane properties}
\label{sec bdy conditions}

The above considerations show that the stability of the linearized
fluctuations is related to the behaviour of the modes as they
approach the branes. In this section we briefly explore to what
extent the properties of the linearized solutions capture what we
know about the asymptotic properties of the exact solutions.
Exploring this connection also allows us to make some contact
between the conditions for stability and the physical properties
of the branes which source the bulk field configurations, through
considerations along the lines of those given in
refs.~\cite{Scaling,GGPplus,NavSant}.

\subsubsection*{Asymptotics in GN gauge}

Recall for these purposes that on general grounds the various bulk
fields are known (in GN gauge) to approach the branes through a
power-law form given by eqs.~\pref{asymptoticfields}. In principle
there are two cases to consider, depending on whether or not the
background and perturbed fields share the same powers in these
asymptotic expressions. That is, if we imagine that the background
and perturbed metric components, $e^{a_0}$ and $e^a = e^{a_0}
e^A$, satisfy
\be
    e^{a_0} \to c_{a0} (H\rho)^{\alpha_0} \quad \hbox{and} \quad
    e^a \to c_a (H\rho)^\alpha \,,
\ee
near $\rho = 0$, then we see that the fluctuation field, $A$, must
satisfy
\be \label{logBC}
    A \to (\alpha - \alpha_0) \, \ln (H\rho) + \ln \left(
    \frac{c_a}{c_{a0}} \right) + \cdots \,,
\ee
in the same limit. Similar expressions also hold for the fields
$B$, $V$, and $\Phi$.

Since the Maxwell field strength varies as $F_{\rho\theta} \propto
\rho^{\gamma + 2\beta} \propto \rho^{p + 2\beta - 1}$ the same
argument has slightly different implications for $\cA_\theta$,
since $a_\theta \propto \rho^{p_0 + 2\beta_0}$ and $A_\theta =
a_\theta + \cA_\theta \propto \rho^{p + 2\beta}$. (Notice that
this gives the expected results $F_{\rho\theta} \propto \rho$ and
$A_\theta \propto \rho^2$ in the conical case, for which $\beta =
1$ and $p=0$.) In this case we have $\cA_\theta \propto \rho^{p +
2\beta}$ if $p+2\beta < p_0 + 2\beta_0$, because the small-$\rho$
behaviour of $A_\theta$ dominates that of $a_\theta$. Things are
different if $p+2\beta > p_0 + 2\beta_0$, however, since in this
case $a_\theta$ dominates $A_\theta$, and in order to achieve this
$\cA_\theta$ must contain a piece which varies in the same way and
cancels the contribution of $a_\theta$ as $\rho \to 0$. As a
consequence in this case $\cA_\theta \propto \rho^{p_0 +
2\beta_0}$ for small $\rho$.

How singular this is depends on the allowed range of $p + 2\beta$.
Notice that in general the constraints $4\alpha + \beta = 4
\alpha^2 + \beta^2 + p^2 = 1$ require $\frac54 \beta^2 - \frac12
\beta + p^2 = \frac34$. Solving this equation for $p$, the sum
$p+2\beta$ is thus minimal (`$-$' sign) and maximal (`$+$' sign)
for $\beta = \frac15 \mp \frac{16}{5\sqrt{21}}$, and $p = \mp
\frac{2}{\sqrt{21}}$, so that the sum is in the range $-1.433
\simeq \frac{2}{5}(1-\sqrt{21}) \le p + 2\beta \le \frac{2}{5}
(1+\sqrt{21}) \simeq 2.233$. We see that perturbations to
$\cA_\theta$ can become quite singular. In the particular case of
conical singularities $p = 0$ and $\beta =1$ and so $p + 2\beta =
2$, leading to smooth behaviour near the branes.

We see from the above that the perturbations need not be smooth at
the brane positions, but that in $GN$ gauge the metric
perturbations can at worst diverge logarithmically in the
near-brane limit. (Perturbations to the Maxwell field can be more
singular than logarithmic, but only if $p$ and $\beta$ are
sufficiently negative.) Since the arguments of
refs.~\cite{Scaling,GGPplus,NavSant} relate powers like $\alpha$
to physical properties on the branes, these logarithmically
singular perturbations only arise if some of the brane properties
are themselves perturbed. It is only in the limit that the
asymptotic near-brane behaviour is the same before and after
perturbation that the perturbed solutions are nonsingular at the
brane positions. This is the case in particular if it is assumed
that the geometry has purely conical singularities before and
after perturbation, or if the sources are represented as delta
functions with the implicit associated assumption that the bulk
fields are well-defined when evaluated at the brane positions --
as is common in higher-codimension stability analyses, such as
that of ref.~\cite{LP}.

\subsubsection*{Asymptotic forms in comoving gauge}

Although we have seen that within GN gauge fluctuations diverge at
most logarithmically near the source branes, we next determine
what this implies for the near-brane behaviour in the (comoving)
gauge we use in our analysis. In this section we examine in
particular the near-brane asymptotics in comoving gauge.

Choosing background proper distance, $\rho$, (rather than $\eta$)
as the radial coordinate we have seen that the perturbations in
comoving and GN gauge are related by eqs.~\pref{com to GN}, which
we reproduce here for ease of reference:
\ba
    &&A^{(c)}= A^{(GN)}+ \varepsilon \partial_\rho a_0 \,,
    \quad
    B^{(c)} = B^{(GN)} + \varepsilon \partial_\rho b_0 \nn\\
    &&\qquad V^{(c)} = \partial_\rho \varepsilon
    \quad\hbox{and}\quad
    \Phi^{(c)}=\Phi^{(GN)} +\varepsilon \partial_\rho
    \varphi_0  \,,
\ea
with $\varepsilon = -\cA_\theta^{(GN)}/(\partial_\rho a_\theta)$.
The condition $M^{(c)} = 0$ is ensured through a change of the
coordinates $x^\mu$ with parameter $\epsilon = -M^{(GN)}/2$.

Since we know how the fluctuations vary with $\rho$ as $\rho \to
0$ within GN gauge, the above equations allow this information to
be carried over to comoving gauge. In particular, since $a_\theta
+ \cA_\theta \propto \rho^{p+2\beta}$ and $a_\theta \propto
\rho^{p_0 + 2\beta_0}$, we saw $\cA_\theta \propto \rho^\zeta$,
where $\zeta = \hbox{min} (p + 2\beta, p_0 + 2\beta_0)$. We see
that $\varepsilon \propto \rho^{1+\Delta}$ and that there are two
cases: ($i$) $\Delta = (p-p_0) + 2(\beta - \beta_0)$ if $p +
2\beta < p_0 + 2\beta_0$, or ($ii$) $\Delta = 0$ if $p + 2\beta
\ge p_0 + 2\beta_0$. Since $a_0$, $b_0$, $\varphi_0$, $A^{(GN)}$,
$B^{(GN)}$ and $\Phi^{(GN)}$ all vary at most logarithmically for
small $\rho$, we see from the above that all of the fluctuations
in comoving gauge are also at worst logarithmic provided only that
$\Delta =  (p + 2\beta) - (p_0 + 2\beta_0 ) \ge 0$. (In
particular, for conical backgrounds the perturbations are at most
logarithmic provided $\Delta = (p + 2\beta - 2) \ge 0$.)

These arguments show that the fluctuations $\Phi$, $B$ and $\Psi$
diverge at most logarithmically within comoving gauge for a broad
range of physical situations. Inspection of the definitions,
eqs.~\pref{eq talpha}--\pref{eq tpsi}, shows that this implies the
same conclusion for $e^{-\cX }\tilde f$, $\tilde\chi$ and
$\tilde\psi$.

\section{General Stability Analysis}
\label{sec stability}

We now give several general arguments in favour of stability for a
broad class of boundary conditions. We provide two types of
arguments, which complement one another. The first of these works
directly with the linearized equations derived above, and can be
used directly with either the longitudinal-gauge (provided in
Appendix \eqref{sec long gauge}) -- or comoving-gauge
formulations. The second argument is instead cast in terms of the
action, and closes a loophole left by the previous
equation-of-motion analysis. We make this second argument only for
comoving gauge (and not for longitudinal gauge, say) due to
complications which arise in constructing the relevant action in
other gauges.

\subsection{Equation of motion and tachyons}
\label{tachyons}
We start by arguing for stability directly with the equations of
motion. The goal of this argument is to relate the sign of the
energy eigenvalue, $\omega^2$, to the boundary conditions which
the fluctuations satisfy near the positions of the source branes.

The most direct way to do so is to multiply eq.~\eqref{eq talpha} by
$\chi^*$; sum the result with its complex conjugate; and
integrate the answer over the extra dimensions, to get:
\be \label{integratedchitilde}
    \omega^2 \int_{-\infty}^{\infty} \, \d \eta\,  e^{2\cY + \cZ}
    \, |\tilde \chi|^2 = -\frac12 \Big[(|\tilde\chi|^2)'\Big]_{-\infty}^{\infty}
    +\int_{-\infty}^{+\infty} \, \d \eta\, | \tilde \chi '|^2 \,.
\ee
This shows how the sign of $\omega^2$ is related to the behaviour
of $\tilde\chi$ near the two brane positions. Notice that the
combination $e^{2\cY + \cZ}$ appearing on the left-hand side of
\pref{integratedchitilde} is related to $\cA$ and $\cW$ by
$e^{2\cy+\cz}=\cA^2\cW^6$.

Although we do not have explicit solutions to the linearized
equations for general non-conical backgrounds, the asymptotic
behaviour of these solutions was argued above to be
the same as in the conical case. In particular, both fluctuations
$\tilde \chi$ and $\tilde \psi$ are typically proportional to
$\log \rho \sim \eta$ as $\eta\rightarrow \pm \infty$, whereas
$\tilde f$ usually diverges as $1/\rho \sim e^{|\eta|}$ in this
limit. Thus the general asymptotic behaviour as $\eta \to \pm
\infty$ is
\ba
    &&\tilde{\chi} \sim \(C_\pm \eta + D_\pm\) \,, \qquad
    \tilde{\psi} \sim \(E_\pm \eta + F_\pm\) \,, \nn \\
    &&\qquad \hbox{ and } \quad \tilde f \sim \(G_\pm\, e^{\lambda_2
    |\eta|}+H_\pm\, e^{- \lambda_2 |\eta|}\) \,,
\ea
for constants $C_\pm$, $D_\pm$, $E_\pm$, $G_\pm$ and $H_\pm$.
Using this in eq.~\pref{integratedchitilde}, and cutting off the
integrations at $\eta = \pm \Lambda$, gives
\be
    \omega^2 \int_{-\Lambda}^{+\Lambda} \, \d \eta\,  e^{2\cY + \cZ}
    \, |\tilde \chi|^2 = -\(|C_+|^2+|C_-|^2\)
    \Lambda -\hbox{Re}\(D_+^*C_+-D_-^* C_-\)
    +\int_{-\Lambda}^{+\Lambda} \, \d \eta\, | \tilde \chi '|^2
    \,.
\ee
We see from this that if the fluctuation is required to remain
finite on both branes --- \ie\ $C_+ = C_- = 0$ --- the boundary
term vanishes and the squared energy of any mode is necessarily
positive: $\omega^2 \ge 0$. Furthermore, $\omega$ only vanishes if
$\tilde \chi$ is a constant throughout the entire bulk.

Applying the same argument to the $\tilde \psi$ equation gives:
\ba
    \omega^2 \int_{-\Lambda}^{+\Lambda} \, \d \eta\, e^{2\cY + \cZ}\,
    |\tilde \psi|^2 &=&-\frac12 \Big[ (|\tilde\psi|^2)'\Big]_{-\Lambda}^{+\Lambda}
    +\int_{-\Lambda}^{+\Lambda} \, \d \eta\, \(| \tilde \psi '|^2+
    \frac{32\,g^2}{\k}\frac{\cV^2 e^{2\cy}}{(2\cy'+\cz')^2}\, |\tilde \psi|^2 \)\nn\\
    &=& -\(|E_+|^2+|E_-|^2\)\Lambda -\hbox{Re} \(F_+^*E_+-F_-^* E_-\) \\
    && \qquad\qquad +\int_{-\Lambda}^{+\Lambda} \, \d \eta\, \(| \tilde \psi '|^2+
    \frac{32\,g^2}{\k}\frac{\cV^2 e^{2\cy}}{(2\cy'+\cz')^2}\, |\tilde \psi|^2
    \).\nn
\ea
Here again, $\omega^2 \ge 0$ if $E_+ = E_- =0$ is imposed so that
the fluctuation remains finite on both branes, and can only vanish
if $\tilde \psi$ vanishes identically.

Finally, the same argument applied
to the mode $\tilde f$ gives
\ba
    \omega^2 \int_{-\Lambda}^{+\Lambda} \, \d \eta\,  e^{2\cY+\cZ}
    \, |\tilde f|^2 &=&-\frac12 \Big[(|\tilde f|^2)'
    \Big]_{-\Lambda}^{+\Lambda}
    +\int_{-\Lambda}^{+\Lambda} \, \d \eta\, \(| \tilde f '|^2+
    \cU^2 \, |\tilde f|^2 \)\nn \\
    &=&-\lambda_2\(|G_+|^2+|G_-|^2\)e^{2\lambda_2 \Lambda}
    +\lambda_2 \(|H_+|^2+|H_-|^2\)e^{-2\lambda_2 \Lambda} \\
    &&\qquad\qquad +\int_{-\Lambda}^{+\Lambda} \, \d \eta\,
    \(| \tilde f '|^2+ \cU^2\, |\tilde f|^2 \),\nn
\ea
so that finiteness of the boundary terms requires $G_\pm = 0$ if
$\lambda_2 > 0$ or $H_\pm = 0$ if $\lambda_2 < 0$. In either case
$\omega^2 \ge 0$ and can only vanish if $\tilde f \equiv 0$
throughout the bulk.

In this way we see that perturbations about any background (let it be
conical or not)   are marginally stable,
provided that the fluctuations are restricted to be nonsingular at
the brane positions.

\subsection{Action analysis and ghosts}
\label{sec action}

The previous argument shows that the eigenmode frequency $\omega$
must be real for a broad choice of boundary conditions at the
brane positions. However one might worry that the Lagrangian
density for the linearized fluctuation might have the form $\cL =
\Phi^* f(\varphi) [\Box - g(\varphi) ] \Phi$, where $\varphi$
(resp. $\Phi$) denotes a generic background (resp. fluctuation) field, and
$f(\varphi)$ and $g(\varphi)$ are background-field dependent
quantities. Notice that the equation of motion for $\Phi$ implies
$[-\Box + g(\varphi)] \Phi = 0$, and so implies $\omega^2 \ge 0$
if $g(\varphi) \ge 0$. But this might still imply a negative
contribution to the fluctuation energy if the prefactor
$f(\varphi)$ should happen to be negative for some configurations
$\varphi$. (Indeed, precisely this form of instability was argued
in ref.~\cite{LP} to occur in 6D chiral supergravity linearized
about rugby-ball configurations due to kinetic-term mixing amongst
various KK modes.) We next present a second argument for
stability, which closes this particular loophole.

The starting point in this approach is to expand the action,
eq.~\eqref{6DSugraAction}, to second order in the perturbations.
Working in comoving gauge and using the coordinate $\eta$ we get
in this way the following quadratic action
\ba
    S^{(2)}&=&\int \d^6
    x\(\mathcal{L}_{\rm{kin}}+\mathcal{L}_{\rm{dyn}}\),\\
    \mathcal{L}_{\rm{kin}}&=&\frac{1}{4\k} e^{\frac 1 2 (\cx -\cy )}
    \Bigl[  6\Psi'-4B ' -4\varphi_0' \Phi
    -2(2\cx '+\cz ') B
    +(2\cy '+\cz ') \xi
    \Bigr] \Box N\\
    && +\frac{3}{4\k}e^{2\cy +\cz }\(2B-\Psi \) \Box \Psi
    +\frac{1}{4\k}e^{2\cy +\cz}(3\Psi-2B)\Box \xi
    +\frac{1}{2\k}e^{2\cy +\cz }\Phi \Box \Phi \nn \\
    \mathcal{L}_{\rm{dyn}}&=&\frac{1}{2\k}\(3\Psi'-4B '\)\Psi'
    +\frac{1}{ \k}(2 \cx '+\cz ')\Psi B '
    -\frac{1}{2\k}\Phi'^2
    -\frac{1}{2\k}(\cy '-\cx ')\xi B ' \\ &&
    -\frac{1}{2\k}\(4\Psi'-2B '+\xi'\)\varphi_0'\Phi
    +\frac{1}{2 \k}(2 \cy '+\cz ')\xi \Psi'
    -\frac{q^2}{4} \, e^{2\cx }\Phi^2 \nn \\&&
    -\frac{q^2}{2} \, e^{2\cx } B  \(B +\xi+4 \Psi+2\Phi\)
    -\frac{g^2}{\kappa^4}e^{2\cy } \(\Phi+2\xi\)\Phi
    -\frac{g^2}{2\kappa^4}e^{2\cy } \xi^2 \,,\nn
\ea
where we allow the fluctuations to depend on all four noncompact
coordinates, $x^\mu$, $\Box = \eta^{\mu\nu} \partial_\mu
\partial_\nu$ denotes the flat d'Alembertian, and the split into
`dyn' and `kin' distinguishes terms which involve $\Box$ from
those which do not. We obtain the above expression by freely
integrating by parts while ignoring the surface terms which this
procedure introduces at the brane positions. This neglect of
surface terms is justified inasmuch as our goal is to exclude the
possibility of the negative (ghost-like) kinetic terms, as
described above.

Differentiating this action with respect to the shift function $N$
leads to the constraint
\ba
    \xi=\frac{2(2\cx'+\cz') B +4\varphi_0'
    \Phi+2\(2B '-3\Psi'\)}{2\cy'+\cz'} \,,
    \label{lapse}
\ea
while differentiating with respect to the lapse function $\xi$
gives the constraint for $N$:
\ba \label{shift}
    \Box N = \frac{e^{\frac 1 2
    \(\cy-\cx\)}}{2\cy'+\cz'} \Bigg[&&\hspace{-7pt} \k q^2 e^{2\cx} \(2B
    +\Phi\)+\frac{4\,g^2}{\k} e^{2 \cy} \(\xi+\Phi\) +e^{2\cy+\cz}\Box
    \(2B -3\Psi\)\\
    &&+2(\cy'-\cx')B '-2(2\cy'+\cz') \Psi'-2\varphi_0' \Phi'
    \Bigg] \,.\nn
\ea These two constraints are consistent with what was obtained in
eqs. \eqref{Eqzeta} and \eqref{EqN}. As before, we may use both
these constraints to obtain the equations of motion for the three
remaining degrees of freedom, $\Phi, B $ and $\Psi$. Substituting
eqs. \eqref{lapse} and \eqref{shift} into the action, together
with the change of variable \eqref{change variable 2}, the
resulting action takes the remarkably simple form
\ba \label{Action decoupled}
    S^{(2)}=\frac{1}{\k} \int \d^6x \Bigg\{
    &&\hspace{-7pt} \tilde \chi \Big[
    \partial_\eta^2 + e^{2\cy+\cz} \Box \Big] \tilde{\chi } +
    \tilde\psi \left[ \partial_\eta^2 + e^{2\cy+\cz}\Box
    -\frac{32\,g^2}{\k}\, \frac{\cV^2 e^{2\cy}}{\(2\cy'+\cz'\)^2}
     \right] \tilde{\psi} \nn\\[-2pt]
    &&\qquad\qquad
    +\tilde{f}\Big[\partial_\eta^2+e^{2\cy+\cz}\Box-\cU^2 \Big]\tilde{f}
    \Bigg\} \,.
\ea
Varying this action gives the same equations as the ones obtained
in (\ref{eq talpha}, \ref{eq tphi}, \ref{eq tpsi}), confirming
that the substitution of ansatz \eqref{com gauge} into the action
consistently reproduces the equations of motion of the full
theory.

Since all of the kinetic terms in the action are positive, it
follows immediately that the theory has no ghost modes (for which
$f(\varphi) < 0$). Because this disagrees with the conclusions
drawn in \cite{LP}, we also reproduce their longitudinal-gauge
analysis in Appendix \ref{appendix LP Comparison} and identify a
sign error which we believe to be the source of the discrepancy.

\section{Conclusions}
\label{sec conclusion}

In this paper we have studied the linearized evolution of
perturbations to gauged chiral 6D supergravity compactified to 4D
on a broad class of static and axially-symmetric vacuum solutions,
sourced by two space-filling 3-branes. We follow previous authors
in focussing on scalar modes which share the axial symmetry of the
background, and which are even under a convenient parity
transformation. These restrictions are consistent with the
equations of motion, since they are enforced by symmetries. They
do not compromise the stability conclusions because the excluded
modes necessarily have higher squared-frequencies than do the
lowest of the modes kept (which we find are bounded below by
$\omega^2 = 0$).

Although none of the backgrounds we perturb are supersymmetric, we
find they are all marginally stable to perturbations which are
well-behaved at the brane positions. The marginal direction is
unique and is the one required on general grounds by a general
scaling property of the 6D supergravity equations. Besides
providing a general stability argument which works for a broad
class of static vacua, we also provide analytic solutions to the
fluctuation equation for the special case of conical backgrounds,
including (but not restricted to) the rugby ball,
(in comoving gauge) and the rugby ball solutions (in longitudinal
gauge). These allow us to explicitly verify that the near-brane
behaviour of the solutions has the properties required by general
asymptotic properties of the field equations. By constructing the
truncated action for the lowest-lying KK modes, we are able to
show that no instabilities arise from mode-mixing amongst the
lowest-lying levels, contrary to recent claims.

\section*{Acknowledgements}

We thank Allan Bayntun, Jonathan Sharmin and Christian Veenstra for
their contributions to preliminary versions of this work. CB would
like to thank the Banff International Research Station
for its kind hospitality while this paper was being finished. AJT
is supported in part by US Department of Energy Grant
DE-FG02-91ER40671. CB and DH acknowledge support from the Natural
Sciences and Engineering Research Council of Canada, with
additional support for CB coming from the Killam Foundation and
McMaster University. DH acknowledges support from les fonds
Qu\'eb\'ecois de la recherche sur la nature et les technologies.
CdR is funded by a grant from the Swiss National Science
Foundation.

\appendix

\section{Analysis in longitudinal gauge}
\label{sec long gauge}
In this section, we linearize the field equations in longitudinal gauge,
specializing to those background configurations for which
$\lambda_3 = 0$ and so which only have conical singularities.
Working in this gauge allows us to make contact with earlier
calculations, and so for ease of comparison in this section we
adopt the notation of these earlier workers. Following
refs.~\cite{Jim,LP} we therefore write
\be \label{long gauge}
    A^{(l)} =  - \frac{\Psi}{2} \,, \quad
    V^{(l)} = \frac{\xi}{2} \,, \quad
    B^{(l)} = \Psi - \frac{\xi}{2} \,, \quad
    \Phi^{(l)} = - \frac{f}{2} \quad \hbox{and} \quad
    \cA_\theta^{(l)} = \cA_\theta \,.
\ee
We also follow refs.~\cite{Jim,LP} and adopt as radial coordinate,
$\rho$, as proper distance in the extra dimensions (measured using
the background metric), leading to the full ansatz
\be \label{metric long gauge}
    \exd s^2 = \cW^2(\rho) \, e^{-\Psi} \,\eta_{\mu\nu} \, \exd
    x^\mu \exd x^\nu + e^{\xi} \, \exd \rho^2 + \cA^2 \, e^{2\Psi
    - \xi} \, \exd \theta^2 \,,
\ee
together with $e^{-2\varphi} = \cW^4 \, e^{f}$ and $A_\theta' =
(q\cA/\cW^6) + \cA_\theta'$. Throughout this appendix we use primes
to denote $\exd/\exd\rho$ rather than $\exd/\exd\eta$.

Fourier transforming in time allows us to write $-\partial_t^2 =
\omega^2$, and so we find that the $(tt)$ and $(ij)$ Einstein
equations degenerate into the same equation at the linearized
level, to give the following $(\mu\nu)$ Einstein equation:
\bea
    &&\frac{\omega^2 \Psi}{\cW^2} + \Psi'' + \left( \frac{6\,\cW'}{\cW} +
    \frac{\cA'}{\cA} \right) \Psi' + \left( \frac{2\,\cW'}{\cW}
    \right) \xi' \\
    &&\qquad\qquad\qquad
    - \frac{\kappa^2}{2} \left( \frac{q^2}{\cW^{10}}
    \right) \left( 2\Psi - \xi - \frac{f}{2} \right) +
    \kappa^2 \left( \frac{ q}{ \cA \cW^4}
    \right) \cA_\theta' - \frac{2\,g^2}{\kappa^2 \cW^2} \left( \xi
    - \frac{f}{2} \right) = 0 \,. \nn
\eea
Similarly, the $(t\rho)$ equation becomes:
\be
    i \omega \left[ \Psi' + \xi'  + 2 \left(\frac{\cW'}{\cW}
    - \frac{\cA'}{\cA} \right) \Psi + 2\left( \frac{\cW'}{\cW} +
    \frac{\cA'}{\cA} \right) \xi - \left( \frac{2\,
    \cW'}{\cW} \right) f - 2 \kappa^2 \left( \frac{q}{
    \cA \cW^4} \right) \cA_\theta \right] = 0 \,.
\ee
The $(\rho\rho)$ Einstein equation becomes
\bea
    &&-\frac{\omega^2 \xi}{2\,\cW^2} + \Psi'' + \frac{\xi''}{2}
    + \left( \frac{4\,\cW'}{\cW}
    -\frac{2\,\cA'}{\cA} \right) \Psi'
    + \frac12\left( \frac{4\,\cW'}{\cW} + \frac{3\, \cA'}{\cA}
    \right) \xi' - \left( \frac{2\, \cW'}{\cW} \right) f' \\
    &&\qquad\qquad\qquad
    +\frac{3\kappa^2}{4} \left( \frac{q^2}{\cW^{10}}
    \right) \left( 2\Psi - \xi - \frac{f}{2} \right) -
    \frac{3\kappa^2}{2} \left( \frac{ q}{ \cA \cW^4}
    \right) \cA_\theta' - \frac{g^2}{\kappa^2 \cW^2} \left( \xi
    - \frac{f}{2} \right) = 0 \,. \nn
\eea
The $(\theta\theta)$ Einstein equation becomes
\bea
    &&-\frac{\omega^2 \Psi}{\cW^2} +\frac{\omega^2 \xi}{2\,\cW^2}
    - \Psi'' + \frac{\xi''}{2} - \left( \frac{4\,\cW'}{\cW} \right) \Psi'
    + \frac12\left( \frac{4\,\cW'}{\cW} + \frac{3\, \cA'}{\cA}
    \right) \xi' \\
    &&\qquad\qquad\qquad
    +\frac{3\kappa^2}{4} \left( \frac{q^2}{\cW^{10}}
    \right) \left( 2\Psi - \xi - \frac{f}{2} \right) -
    \frac{3\kappa^2}{2} \left( \frac{ q}{ \cA \cW^4}
    \right) \cA_\theta' - \frac{g^2}{\kappa^2 \cW^2} \left( \xi
    - \frac{f}{2} \right) = 0 \,. \nn
\eea
The dilaton equation becomes
\bea
    &&\frac{\omega^2 f}{\cW^2} + f''
    - \left( \frac{4\,\cW'}{\cW} \right) \Bigl( \Psi'
    + \xi' \Bigr) + \left( \frac{4\,\cW'}{\cW} + \frac{\cA'}{\cA}
    \right) f'  \\
    &&\qquad\qquad\qquad
    +\kappa^2 \left( \frac{q^2}{\cW^{10}}
    \right) \left( 2\Psi - \xi - \frac{f}{2} \right) -
    2\kappa^2 \left( \frac{ q}{ \cA \cW^4}
    \right) \cA_\theta' + \frac{4\,g^2}{\kappa^2 \cW^2} \left( \xi
    - \frac{f}{2} \right) = 0 \,, \nn
\eea
and the Maxwell equation becomes
\be
    \frac{\omega^2 \cA_\theta}{\cW^2} + \cA_\theta''
    + \left( \frac{6\,\cW'}{\cW} - \frac{\cA'}{\cA}
    \right) \cA_\theta' - \left( \frac{  q \cA}{\cW^6}
    \right) \left( 3 \Psi' - \frac{f'}{2} \right) = 0 \,.
\ee

Nominally we have here 6 equations --- dilaton, Maxwell, and the
($tt$), $(t\rho)$, $(\rho\rho)$ and $(\theta\theta)$ Einstein
equations --- for the 4 unknown functions, $\Psi$, $\xi$, $f$ and
$\cA_\theta$. However, as was the case in comoving gauge, the
Bianchi identities ensure that two of these equations are not
independent. This is because two combinations of these equations
can be chosen to only depend on single derivatives with respect to
time, and the Bianchi identities ensure that the solutions to
these lower-order constraint equations is consistent with the time
evolution. This allows us to use the constraint equations to solve
for two of the fields, and then to drop two of the other field
equations as redundant.

Following this logic we determine $\cA_\theta$ by using the
$(t\rho)$ Einstein equation, which integrates to
\be \label{SolnForA}
    \left( \frac{\kappa^2 q}{ \cA \cW^4} \right) \cA_\theta =
    \frac12 (\Psi' + \xi') +  \left( \frac{\cW'}{\cW}
    - \frac{\cA'}{\cA} \right) \Psi + \left( \frac{\cW'}{\cW} +
    \frac{\cA'}{\cA} \right) \xi - \left(
    \frac{\cW'}{\cW} \right) f + \Omega(\rho) \,,
\ee
where $\Omega(\rho)$ is an arbitrary function of integration. This
unknown function can be determined by using the equation obtained
by adding the $(tt) + (\rho\rho) + (\theta\theta)$ Einstein
equations (which is a constraint inasmuch as it does not contain
any time derivatives):
\bea \label{LPConstraint}
    &&\Psi'' + \xi'' + \left( \frac{6\, \cW'}{\cW} -
    \frac{\cA'}{\cA} \right) \Psi' + \left( \frac{6\, \cW'}{\cW}
    + \frac{3\, \cA'}{\cA} \right) \xi' - \left( \frac{2 \cW'}{\cW}
    \right) f' \\
    &&\qquad\qquad + \left( \frac{\kappa^2 q^2}{\cW^{10}} \right)
    \left( 2 \Psi - \xi - \frac{f}{2} \right) - \left( \frac{2
    \kappa^2 q}{ \cA \cW^4} \right) \cA'_\theta - \frac{4
    g^2}{\kappa^2 \cW^2} \left( \xi - \frac{f}{2} \right)
    = 0 \,. \nn
\eea
Substituting eq.~\pref{SolnForA} into eq.~\pref{LPConstraint} and
using the background equations of motion leads to the following
condition for $\Omega$:
\be
    (\cA \cW^4 \Omega)' = 0\, ,
\ee
and using this in the derivative of eq.~\pref{SolnForA} gives
\bea \label{SolnForAPrime}
    &&\left( \frac{\kappa^2 q}{ \cA \cW^4} \right) \cA'_\theta =
    \frac12 (\Psi'' + \xi'') + \left( \frac{3\, \cW'}{\cW}
    - \frac{\cA'}{2\,\cA} \right) \Psi' + \left( \frac{3\,\cW'}{\cW}
    + \frac{3\,\cA'}{2\,\cA} \right) \xi'  - \left(
    \frac{\cW'}{\cW} \right) f' \nn\\
    &&\qquad\qquad\qquad\qquad
    + \frac12 \left( \frac{\kappa^2 q^2}{\cW^{10}} \right)
    \left( 2 \Psi - \xi - \frac{f}{2} \right)  - \frac{2
    g^2}{\kappa^2 \cW^2} \left( \xi - \frac{f}{2} \right)
    \,.
\eea

We may now use this last expression to eliminate $\cA'_\theta$
from the remaining 3 field equations, which we take to be the
dilaton and $(tt)$ and $(\rho\rho)$ Einstein equations, leading
to:
\be \label{SolnForPsiZetaprimeprime}
    \frac{\omega^2 \Psi}{\cW^2} + \frac32 \Psi'' +
    \frac12 \,\xi'' + \left(\frac{9\,\cW'}{\cW} + \frac{\cA'}{2\, \cA}
    \right) \Psi' + \left( \frac{5\,\cW'}{\cW} +
    \frac{3\, \cA'}{2\, \cA} \right) \xi'
    - \left( \frac{\cW'}{\cW} \right) f'
    - \frac{4\,g^2}{\kappa^2 \cW^2} \left( \xi
    - \frac{f}{2} \right) = 0 \,,
\ee
\be \label{SolnForPsiZeta2primeprime}
    \frac{\omega^2 \xi}{\cW^2} - \frac12 \Psi'' + \frac12 \,\xi''
    + \left( \frac{\cW'}{\cW} + \frac{5\,\cA'}{2\,\cA}
    \right) \Psi' + \left( \frac{5\,\cW'}{\cW} + \frac{3\,\cA'}{2\,\cA}
    \right) \xi' + \left( \frac{\cW'}{\cW} \right) f' -
    \frac{4\,g^2}{\kappa^2 \cW^2} \left( \xi
    - \frac{f}{2} \right) = 0 \,,
\ee
and
\bea \label{SolnForfprimeprime}
    &&\frac{\omega^2 f}{\cW^2} + f'' - \Psi'' - \xi''
    - \left( \frac{10\,\cW'}{\cW} - \frac{\cA'}{\cA}
    \right) \Psi' - \left( \frac{10\,\cW'}{\cW} + \frac{3\,\cA'}{\cA}
    \right) \xi' \\
    &&\qquad\qquad\qquad\qquad\qquad\qquad\qquad\qquad\qquad
    + \left( \frac{6\,\cW'}{\cW} + \frac{\cA'}{\cA}
    \right) f' + \frac{8\,g^2}{\kappa^2 \cW^2} \left( \xi
    - \frac{f}{2} \right) = 0 \,, \nn
\eea
in agreement with eqs.~(37)--(39) of ref.~\cite{LP}. These
equations easily lend themselves to numerical integration because
they are independent, since all of the constraints have been
explicitly solved.

The above equations take a somewhat simpler form if rewritten in
terms of new variables: $\chi = 2\Psi + f = - 4A - 2\Phi =
2(B+V-\Phi)$ and $\Gamma = \Psi + \xi = 2(V-A) = B + 3V$:
\be \label{chiEqn}
    \frac{\omega^2 \chi}{\cW^2} + \chi''
    + \left(\frac{4\,\cW'}{\cW} + \frac{\cA'}{\cA}
    \right) \chi' = 0 \,,
\ee
\be \label{GammaEqn}
    \frac{\omega^2 \Gamma}{\cW^2} + \Gamma''
    + \left( \frac{10\,\cW'}{\cW} + \frac{3\,\cA'}{\cA}
    \right) \Gamma' - \left( \frac{8\,g^2}{\kappa^2 \cW^2}
    \right) \Gamma  = - \left( \frac{4\,g^2}{\kappa^2 \cW^2}
    \right)  \chi \,,
\ee
and
\be \label{NewfEqn}
    \frac{\omega^2 f}{\cW^2} + f'' + \left( \frac{6\,\cW'}{\cW} - \frac{\cA'}{\cA}
    \right) f' = - \frac{\omega^2 \Gamma}{\cW^2} - \left(
    \frac{2\,\cA'}{\cA} \right) \chi'\,.
\ee
The logic to solving these is to integrate eq.~\pref{chiEqn} for
$\chi$, and then to use the result as a source in
eqs.~\pref{GammaEqn} and \pref{NewfEqn} to integrate these. These
equations may be solved in closed form for the case of the
rugby-ball solutions, as we show in what follows.

Although we arrive at equations which are in perfect agreement
with ref.~\cite{LP}, we shall differ in the conclusion we draw
from them inasmuch as we find the system to be marginally stable
(modulo an issue concerning boundary conditions which is
irrelevant for the comparison with \cite{LP}). For this reason we
also review the stability analysis of ref.~\cite{LP} in appendix
\ref{appendix LP Comparison} and explain the origin of the
discrepancy.

\subsection{Analytic Solutions for the Rugby Ball}

The rugby ball is the special conical geometry for which $e^{w_0}
= e^{a_0} = \cW = e^{\varphi_0} = 1$. In this case explicit forms
for the axially-symmetric scalar mode functions can be found
analytically in longitudinal gauge in terms of well-known special
functions. This has the advantage of providing a concrete verification
of the previous stability arguments, as well as providing
a framework within which to see how the boundary
conditions give rise to the quantization of Kaluza-Klein
frequencies.

Recalling that in this section, we use the proper distance, $\rho$, as radial coordinate the
rugby-ball geometry is given by
\be
    e^{b_0} = \cA = \lambda \, r_0 \sin
    \left( \frac{\rho}{r_0} \right)
    \quad \hbox{where} \quad
    r_0 = \frac{\kappa}{2g} \,.
\ee
These choices imply $\lambda = 1-\delta$ encodes the defect angle,
$2\pi \delta$, at the branes situated at $\rho = 0$ and $\rho =
\pi r_0$.

Using the rugby-ball conditions $\cW' =0$ and $\cA'/\cA = r_0^{-1}
\cot(\rho/r_0)$ allows the fluctuation equations in longitudinal
gauge, eqs.~\pref{chiEqn}--\pref{NewfEqn}, to be simplified to
\bea \label{EqnsRB}
    \omega^2 \chi + \chi'' + \frac{\cot(\rho/r_0)}{r_0} \, \chi' &=& 0
    \nn\\
    \omega^2 \Gamma + \Gamma''
    + \frac{3\cot(\rho/r_0)}{r_0} \,
    \Gamma' - \frac{2\, \Gamma}{r_0^2}
      &=& - \frac{\chi}{r_0^2}  \\
    \omega^2 f + f''  - \frac{\cot(\rho/r_0)}{r_0} \,  f' &=& - \omega^2 \Gamma -
    \frac{2\cot(\rho/r_0)}{r_0} \, \chi'\,. \nn
\eea
To solve these equations
we write $\omega^2 = \mu^2/r_0^2$ and change variables to $z =
\cos(\rho/r_0)$, leading to a relatively familiar set of equations
\bea \label{EqnsRBz}
    \frac{\exd^2 \chi}{\exd z^2} - \left( \frac{ 2z}{1 - z^2}
    \right) \frac{\exd \chi}{\exd z} + \left( \frac{\mu^2}{1 -
    z^2} \right) \chi &=& 0
    \nn\\
    \frac{\exd^2 \Gamma}{\exd z^2} - \left( \frac{ 4z}{1 - z^2}
    \right) \frac{\exd \Gamma}{\exd z} + \left( \frac{\mu^2-2}{1 -
    z^2} \right) \Gamma
      &=& - \frac{\chi}{1-z^2}  \\
    \frac{\exd^2 f}{\exd z^2}  + \left( \frac{\mu^2}{1 -
    z^2} \right) f
      &=& - \left( \frac{\mu^2}{1-z^2} \right) \Gamma
    + \left( \frac{ 2z}{1 - z^2}
    \right) \frac{\exd \chi}{\exd z}\,.  \nn
\eea

In terms of the variable $z = \cos(\rho/r_0)$ the brane
singularities correspond to $z = \pm 1$, at which points we have
\bea
    z &=& 1 - \frac{\rho^2}{2} + \cdots \,, \quad \hbox{near $\rho
    = 0$} \nn\\
    z &=& -1 + \frac{(\rho- \pi r_0)^2}{2} + \cdots \,, \quad \hbox{near $\rho
    = \pi r_0$} \,,
\eea
and so $\rho - \rho_\pm = \mp \sqrt{2} (1 \pm z)^{1/2} + \cdots$ where $\rho_+
= \pi r_0$ and $\rho_- = 0$. The asymptotic near-brane boundary
conditions are therefore related to the behaviour of the solutions
as $z \to \pm 1$.
The asymptotic form which is appropriate to longitudinal gauge may
be found by performing the transformation from GN gauge, along the
lines as was done in the main text for comoving gauge. In this section we concentrate on finding those
solutions for which all perturbations are less singular than
$(1-z^2)^{-1} \propto \rho^{-2}$, including in particular those
for which the divergence at $z \to \pm 1$ is at most logarithmic.

\subsubsection*{Homogeneous solutions}

We start by providing explicit integrals of the homogeneous
parts of eqs.~\pref{EqnsRBz}, for which the right-hand sides
vanish. The first of these is the Legendre equation and therefore
has as solutions
\be \label{ChiLegendre}
    \chi(z) = C_1 \, P_\nu(z) + C_2 \, \hbox{Re} \, Q_\nu(z) \,,
\ee
where $\nu (\nu + 1) = \mu^2$, or $\nu = \nu_\pm = \frac12[ -1 \pm
(1 + 4\mu^2)^{1/2}]$. We take the real part of $Q_\nu$ here
because this function is normally complex for $|z| < 1$. (See
Appendix \ref{app special functions} for our detailed
conventions.) Since $\nu_- = -1 - \nu_+$ and the differential
equation is invariant under $\nu \to -1 -\nu$ it suffices in what
follows to use only the positive root, $\nu = \frac12[ \sqrt{1 + 4
\mu^2} - 1]$.

For generic $\nu$ both $P_\nu$ and $Q_\nu$ have logarithmic
singularities as $z \to \pm 1$. However, we shall find that for
generic $\nu$, $\Gamma$ diverges like $(1\pm z)^{-1}$. Restricting
to at most logarithmic solutions therefore requires us to take
$\nu = \ell = 0,1,2,\dots$, and so we only consider these values
from this point on. (However, we emphasize that since logarithmic
singularities are physically acceptable for the present problem,
this quantization is {\it not} required by the boundary conditions
for $\chi$.) The mass formula
\be \label{massgapformula}
    \omega^2 = \frac{\ell(\ell + 1)}{r_0^2}
    = \frac{4g^2 }{\kappa^2} \, \ell(\ell +1)\,,
\ee
then shows that stability directly follows from this quantization,
in agreement with the general arguments of earlier sections.

Given $\nu = \ell$ the hypergeometric series for $P_\ell(z)$
terminates and degenerates to the Legendre polynomials, which are
regular at {\it both} $z = \pm 1$. Nevertheless, since $Q_\nu(z)$
has the asymptotic limit
\be
    Q_\nu(z) =
    \pm\frac{1}{2\(\pm1\)^\nu}
    \Bigl[ - \ln \left( 1 - z^{-2} \right)+O(1) \Bigr] \quad
    \hbox{as $z \to \pm 1$} \,, \nn\\
\ee
it clearly retains its logarithmic behaviour at $z \to \pm 1$ even
when $\nu = \ell$. Combining the above expressions allows the
near-brane singularities to be written in terms of the remaining
integration constants, $C_1$ and $C_2$, as follows:
\bea
%
    \chi(z) &=& - \, \frac{C_2} 2 \,  \ln \(1-z\)+O(1)
    \qquad\quad\;
    \hbox{as $z \to 1$} \nn\\
    &=&  \frac{C_2}{2(-1)^\ell} \, \ln \(1+z\)+O(1)
    \qquad \hbox{as $z \to -1$} \,.
\eea

The homogeneous part of the equation for $\Gamma$ is also of
Hypergeometric form, leading to the following general homogeneous
solutions:
\be
    \Gamma_h(z) = C_3 F \left[ 1 + \frac{\nu}{2} \,, \frac12 -
    \frac{\nu}{2} \,; \frac12 ; {z}^{2} \right]
    + C_4 \, z F \left[ \frac32 + \frac{\nu}{2} \,,
    1 - \frac{\nu}{2} \,; \frac32 ;{z}^{2} \right]  \,.
\ee
Notice that the function multiplied by $C_3$ (or by $C_4$) is even
(or odd) under reflections about the `equator' at $\rho = \pi
r_0/2$, since these correspond to $z \to - z$. In the special case
$\nu = 0$ the Hypergeometric series can be summed explicitly to
give
\be
    F[1,b;b;z^2] = \sum_{k = 0}^\infty z^{2k}  = \frac{1}{1-z^2}\, ,
\ee
showing that in this case the two homogeneous solutions are the
elementary functions $(1-z^2)^{-1}$ and $z (1-z^2)^{-1}$.

The asymptotic behaviour of the Hypergeometric functions as $z \to
\pm 1$ turn out to be
\bea
    F\left[ 1 + \frac{\nu}{2} \,, \frac12 - \frac{\nu}{2} \,;
    \frac12\,; z^2 \right] &=&
    \frac{\sqrt\pi}{\Gamma\left(1+\frac{\nu}{2}\right) \Gamma\left(
    \frac12 - \frac{\nu}{2} \right)} \left[ \frac{1}{1-z^2} -
    \frac{\nu}{2} \left( \frac12 + \frac{\nu}{2} \right) \ln (1 - z^2) +
    O(1) \right] \,, \nn\\
    F\left[ 1 - \frac{\nu}{2} \,, \frac32 + \frac{\nu}{2} \,;
    \frac32 \,; z^2 \right] &=&  \frac{\sqrt\pi}{2\, \Gamma
    \left(1-\frac{\nu}{2}\right) \Gamma\left(
    \frac32 + \frac{\nu}{2} \right)} \left[ \frac{1}{1-z^2} -
    \frac{\nu}{2} \left( \frac12 + \frac{\nu}{2} \right) \ln (1 - z^2) +
    O(1) \right] \,, \nn\\
\eea
which shows in particular that $\Gamma_h$ diverges like
$(1-z^2)^{-1}$ for generic $\nu$. Since we seek solutions which
only diverge logarithmically, we must choose either $C_3$ or $C_4$
to vanish, and then choose $\nu$ so that the Hypergeometric series
for the other solution must terminate. That is, either $C_4 = 0$
and $\frac12 - \frac{\nu}{2} = -n$ (and so $\nu = 1 + 2n$) or $C_3
= 0$ and $1 - \frac{\nu}{2} = -n$ (and so $\nu = 2(n+1)$), where
$n = 0,1,2,...$ is a non-negative integer. These are automatically
satisfied for non-negative integers, $\nu = \ell \ne 0$, with even
(odd) $\ell$ corresponding to $C_3 = 0$ ($C_4 = 0$). For $\nu = 0$
we have seen that the homogeneous solution is $(C_3 + C_4 \,
z)/(1-z^2)$, which is only nonsingular at both $z = 1$ and $z=-1$
if $C_3 = C_4 = 0$. We see in this way why it is that $\nu$ must
be quantized to be a non-negative integer.

The nonsingular homogeneous solutions for $\Gamma_h$ for the first
few choices for $\ell$ are:
\bea
    \ell = 0 \quad &&\hbox{implies} \quad
    \Gamma_h = 0 \,; \nn\\
    \ell = 1 \quad &&\hbox{implies} \quad
    \Gamma_h = C_3 F \left[ \frac32, 0 ; \frac12 ; z^2 \right]
    = C_3 \,;\nn\\
    \ell = 2 \quad &&\hbox{implies} \quad
    \Gamma_h = C_4 \, z F \left[ \frac52, 0 ; \frac32; z^2 \right]
    = C_4  \, z  \,; \nn\\
    \ell = 3 \quad &&\hbox{implies} \quad
    \Gamma_h = C_3 F \left[ \frac52, -1; \frac12; z^2 \right]
    = C_3(1 - 5 z^2)  \,; \nn
\eea
and so on.

The homogeneous part of the equation for $f$ is also
Hypergeometric, and is given in this case by
\be
    f_h(z) = C_5 (1-z^2) F \left[ 1 + \frac{\nu}{2} \,, \frac12 -
    \frac{\nu}{2} \,; \frac12 ; {z}^{2} \right]
    + C_6 \, z (1-z^2) F \left[ \frac32 + \frac{\nu}{2} \,,
    1 - \frac{\nu}{2} \,; \frac32 ;{z}^{2} \right]  \,.
\ee
Notice that these are the same Hypergeometric functions as appear
in $\Gamma$, and so their singular properties can be read off
from the expressions given above for $\Gamma_h$. In particular,
because of the additional factor of $(1-z^2)$ the asymptotic forms
do not diverge as $z \to \pm 1$ even if the Hypergeometric series
does not terminate, leading to the limits
\bea
    (1-z^2) F\left[ 1 + \frac{\nu}{2} \,, \frac12 - \frac{\nu}{2} \,;
    \frac12\,; z^2 \right] &\to&
    \frac{\sqrt\pi}{\Gamma\left(1+\frac{\nu}{2}\right) \Gamma\left(
    \frac12 - \frac{\nu}{2} \right)} \,, \nn\\
    (1-z^2) F\left[ 1 - \frac{\nu}{2} \,, \frac32 + \frac{\nu}{2} \,;
    \frac32 \,; z^2 \right] &\to&  \frac{\sqrt\pi}{2\,
    \Gamma\left(1-\frac{\nu}{2}\right) \Gamma\left(
    \frac32 + \frac{\nu}{2} \right)} \,.
\eea
For the first few choices for $\ell$ this leads to the following
homogeneous solutions
\bea
    \ell = 0 \quad \hbox{implies} \quad
    f_h &=& C_5 + C_6 \, z \,; \nn\\
    \ell = 1 \quad \hbox{implies} \quad
    f_h &=& (1-z^2) \left( C_5  + C_6 \, z F \left[ 2, \frac12
    ; \frac32 ; z^2 \right]
    \right) \,;\nn\\
    &=& C_5 (1-z^2) - \frac{C_6\, z}{4} \left[ (1-z^2)
    \ln\left( \frac{1+z}{1-z} \right) - z^2 \right] \nn\\
    \ell = 2 \quad \hbox{implies} \quad
    f_h &=& (1-z^2) \left( C_5 \, F \left[ 2, -\frac12 ; \frac12 ; z^2 \right]
    + C_6 \, z \right) \,; \nn\\
    &=& C_5 \left[ \frac{3\,z}{4} (1-z^2) \ln \left( \frac{1+z}{1-z}
    \right) + \frac{3\,z^2}{2} - 1 \right] + C_6 z(1 - z^2) \nn\\
    \ell = 3 \quad \hbox{implies} \quad
    f_h &=& (1-z^2) \left( C_5 (1 - 5 z^2) + C_6 \, z
    F \left[ 3,- \frac12 ; \frac32 ; z^2 \right]
    \right) \,; \nn
\eea
and so on. We see that because of the quantization, $\nu = \ell$,
the only logarithmic singularities in the solutions are those
appearing through $Q_\ell(z)$ in $\chi$.

\subsubsection*{Perturbing tensions}


Dividing the perturbations into those which are even or odd under
the reflection $z \to -z$ allows an even more precise
interpretation.
Finite perturbations which are even under this
reflection correspond to perturbations which change the tension of
both boundary branes in the same way, and so remain within the
class of rugby-ball solutions (for which both brane tensions must
be equal). Conversely, regular perturbations which are odd under $z \to -z$ correspond to
perturbations which change the two brane tensions oppositely, and
so describe excursions into the more general class of warped,
conical solutions \cite{SLED2,GGP,GGPplus}. Logarithmically diverging
perturbations describe the transition from the conical to nonconical
class of solutions.

\subsubsection*{Particular integrals}

We now return to the problem of solving the differential
equations for $\Gamma$ and $f$, including now the right-hand sides
of these equations. Since the general solution is found by adding
the general solution to the homogeneous equation to any particular
solution, it suffices here to identify particular integrals to
these differential equations, which we denote by $\Gamma_{pi}(z)$
and $f_{pi}(z)$. The general solutions are then simply $\Gamma(z)
= \Gamma_h(z) + \Gamma_{pi}(z)$ and $f(z) = f_h(z) + f_{pi}(z)$.

For instance, if we specialize to nonsingular solutions at both $z
= \pm 1$ we may choose $C_2 = 0$ and so take $\chi$ to be given by
the lowest few Legendre polynomials. This leads to the following
expressions for $\Gamma(z) = \Gamma_h(z) + \Gamma_{pi}(z)$:
\bea
    \ell = 0 \quad &&\hbox{implies} \quad
    \Gamma = \Gamma_{pi}(z) = \frac{C_1}{2} \,; \nn\\
    \ell = 1 \quad &&\hbox{implies} \quad
    \Gamma = C_3 F \left[ \frac34, 0 ; \frac12 ; z^2 \right] + \Gamma_{pi}(z)
    = C_3 + \frac{C_1 z}{4} \,;\\
    \ell = 2 \quad &&\hbox{implies} \quad
    \Gamma = C_4 \, z F \left[ \frac52, 0 ; \frac32; z^2 \right]
    + \Gamma_{pi}(z)
    = C_4 \, z - \frac{C_1 z^2}{4}   \,;  \nn
\eea
and so on.

With these solutions in hand we next find the relevant particular
integrals for the differential equation for $f$. Again restricting
to nonsingular solutions, specializing to the first few Legendre
polynomials, and summing $f = f_h + f_{pi}$ gives:
\bea
    &&\ell = 0  \quad \hbox{implies} \quad
    f(z) = C_5 + C_6 z \,;
    \nn\\
    &&\ell = 1  \quad \hbox{implies} \quad
    f(z) = (1-z^2) \left( C_5
    + C_6 \, z F \left[ 2, \frac12 ; \frac32 ; z^2 \right]
    \right) - C_3 + \frac{3\,C_1z}{4} \,;
    \\
    &&\ell =2  \quad \hbox{implies} \quad
    f(z) = (1-z^2) \left( C_5 \, F \left[ 2, -\frac12 ; \frac12 ; z^2 \right]
    + C_6 \, z \right) + \frac{5\,C_1}{8} \Bigl( 3\, z^2 -1 \Bigr)
    -  C_4 \, z  \,;
    \nn
\eea
and so on for as many modes as are desired.

A similar process can be undergone in the more general case where
$C_2 \ne 0$, and so $\chi = C_1 \, P_\ell(z) + C_2 \, \hbox{Re} \,
Q_\ell(z)$. In this case $\chi$ diverges logarithmically in the
near-brane limit, and the particular integrals are no longer
obtainable in such a simple closed form. Nonetheless they may be
obtained numerically in terms of integrals over the
right-hand-side of the equations weighted by appropriate
combinations of Hypergeometric functions.

\subsection{Stability Analysis}

Using the explicit previous solutions and particularly their
asymptotic behaviour, we give here a general argument in favour of
stability for a broad class of boundary conditions.

As performed in section \ref{tachyons}, we argue for stability directly with the equations of
motion and relate the sign of the
energy eigenvalue, $\omega^2$, to the boundary conditions which
the fluctuations satisfy near the positions of the source branes.

Proceeding as before, we multiply eq.~\pref{chiEqn} by $
\cA \cW^4\chi^*$; sum the result with its complex conjugate; and
integrate the answer over the extra dimensions, to get:

\bea \label{chiStable}
    \omega^2 \int_{0}^{\rho_1} \exd \rho \, \cA \cW^2 \;|\chi|^2 &=&
    - \frac12 \int_{0}^{\rho_1} \exd \rho \; \Bigl\{ \cA \cW^4 [
    \chi^* \chi'' + (\chi^*)'' \chi ] + (\cA \cW^4)' (|\chi|^2)' \Bigr\} \nn\\
    &=& - \left[ \frac12 (\cA \cW^4) (|\chi|^2)' \right]^{\rho_1}_{0}
    + \int_{0}^{\rho_1} \exd \rho \; \cA \cW^4 |\chi'|^2  \,.
\eea
We denote here the brane positions by $\rho = 0$
and $\rho = \rho_1$. Since $\sqrt{-g} = \cA \cW^4
> 0$ we see that $\omega^2 \ge 0$ follows provided we can show
that $\chi$ satisfies boundary conditions for which the
combination $\cA\cW^4 \exd(|\chi|^2)/\exd\rho$ vanishes at the
brane positions.

In general the appropriate boundary conditions in longitudinal
gauge for $\chi$ have $\chi \to (H\rho)^{\Delta_l}$ in the
near-brane limit, $\rho \to 0$, with $\Delta_l$ being a power
which is determined in terms of the asymptotics of the background
configuration, and $\Delta_l = 0$ could correspond to
perturbations which remain finite at the branes as well as those
which diverge only logarithmically at the brane positions. Using
this asymptotic limit we have
\be
    (\cA\cW^4) \frac{\exd (|\chi|^2)}{\exd\rho}
    \propto \rho \, \frac{ \exd (|\chi^2|)}{\exd\rho}
    \propto \rho^{2\Delta_l} \,,
\ee
as $\rho \to 0$. This shows that the integral on the left-hand
side generically diverges due to its singularity near the branes,
unless the combination $\rho\, \exd(|\chi|^2)/\exd\rho$ is well
defined at $\rho = 0$. For the brane at $\rho = \rho_1$ the same
argument may be repeated using the coordinate $\hat\rho = \rho_1 -
\rho$, which increases as one moves away from $\hat\rho = 0$.
Keeping in mind that $\exd\hat\rho = - \exd\rho$ but $\sqrt{-g}
\propto +\hat\rho$ as $\hat\rho \to 0$, we see that
\be
    -\left[ \cA\cW^4 \, \exd(|\chi|^2)/\exd \rho \right]^{\rho_1}_0 =
    + \left[\hat\rho \, \exd(|\chi|^2) / \exd\hat\rho \right]_{\hat\rho = 0}
    + \left[ \rho \, \exd(|\chi|^2) / \exd\rho \right]_{\rho = 0} \,,
\ee
and so in the special case that these limits are nonsingular at
the brane position (such as is assumed when the brane is
represented as a delta-function source, or if both the background
and perturbed geometries have only conical singularities) we see
that $\omega^2 \ge 0$ provided that $(|\chi|^2)' \ge 0$ in the
near-brane limit. Furthermore, in this case $\omega = 0$ if and
only if $\chi' = 0$.

Since the above argument assumes that the eigenmode of interest
has $\chi \ne 0$, the only remaining step which is required to
conclude that the system is marginally stable is to separately
show that $\omega^2 \ge 0$ in the special case where $\chi = 0$.
In this case we may apply the same reasoning to
eq.~\pref{GammaEqn}, after multiplying through by $\cA^3\cW^{10}
\Gamma^*$, to get
\be
    \omega^2 \int_{0}^{\rho_1}\exd\rho \, \cA^3 \cW^8
    |\Gamma|^2 = - \left[ \frac12 (\cA^3 \cW^{10}) (|\Gamma|^2)'
    \right]^{\rho_1}_{0} + \int_{0}^{\rho_1} \exd \rho \;
    \left[ (\cA^3 \cW^{10}) |\Gamma'|^2 + \frac{8\,g^2}{\kappa^2} \,
    (\cA^3 \cW^8) |\Gamma|^2 \right] \,.
\ee
The boundary condition can be examined as above, with $\cA^3
\cW^{10} = (\cA \cW^4)^3/\cW^2 \propto \rho^{3-2\alpha_0}$. Since
$\alpha_0 = 0$ for conical backgrounds this vanishes as $\rho^3$.
We see again that $\omega^2 \ge 0$ if $\rho^{3} \exd( |\Gamma|^2)
/ \exd\rho$ is well-defined and non-negative at the brane
position, $\rho \to 0$. In this case $\omega = 0$ implies $\Gamma'
= \Gamma = 0$.

This leaves open only the special case of modes for which both
$\chi$ and $\Gamma$ vanish, and for these modes we repeat the
reasoning by multiplying eq.~\pref{NewfEqn} through by
$(\cW^6/\cA) f^*$, leading to
\be
    \omega^2 \int_{0}^{\rho_1} \exd\rho \, \left(
    \frac{\cW^4}{\cA} \right)
    |f|^2 = - \left[ \frac12 \left( \frac{\cW^6}{\cA} \right) (|f|^2)'
    \right]^{\rho_1}_{0} + \int_{0}^{\rho_1} \exd \rho \;
    \left( \frac{\cW^6}{\cA} \right) |f'|^2  \,.
\ee
In this case $\cW^6/\cA =(\cA\cW^4)(\cW/\cA)^2 \propto
\rho^{1+2(\alpha_0-\beta_0)} \propto \rho^{(3-5\beta_0)/2}$. Since
$\beta_0 = 1$ for conical backgrounds this varies as $\rho^{-1}$
as $\rho \to 0$, so if $\rho^{-1} \, \exd(|f|^2)/\exd\rho$ is well
defined and non-negative at the brane position, $\rho \to 0$, we
find $\omega^2 \ge 0$ with $\omega = 0$ if and only if $f' = 0$.

In this way we see how conclusions about the stability of the
system can be related to asymptotic near-brane behaviour. Although
the argument is inconclusive when the near-brane limit of the
fluctuations is too singular, we do see that when the bulk
fluctuations are nonsingular at the brane positions then the
system is marginally stable. The stability is only marginal
because of the known zero mode, for which $\Gamma =\chi= 0$ while $f$ is a nonzero constant.

\section{Comparison with previous work}
\label{appendix LP Comparison}

The previous stability argument, together with the one performed in
comoving gauge in section \ref{sec stability}, seems in contradiction with the results of
ref.\cite{LP}. We therefore briefly review their analysis in this
section and pin down the reason of the discrepancy.

To analyze the system of linearized equations (\ref{NewfEqn}, \ref{GammaEqn}, \ref{chiEqn}),
ref.~\cite{LP} starts from the assumption
that all modes are proportional to the curvature perturbation
$\Psi$ (see eq.~(41) of \cite{LP}). Using this assumption, they
find that the only existing modes are of the form
\ba
    \(\begin{array}{c}
    f=2\Psi \\
    \Gamma=2\Psi\\
    \chi=4 \Psi
    \end{array}\), \hspace{20pt}
    \(\begin{array}{c}
    f=-2\Psi\\
    \Gamma=2\Psi\\
    \chi=0
    \end{array}\) \hspace{10pt}\rm{and}\hspace{10pt}
    \(\begin{array}{c}
    \frac{\omega^2}{\cW^2} \Psi+ \Psi'' + \left( \frac{6\,\cW'}{\cW} - \frac{\cA'}{\cA}
    \right) \Psi'=0\\
    \Gamma=0\\
    \chi=0
    \end{array}\),
\ea
from which they conclude that ``the spectrum of the scalar
excitations consists of a zero mode, a first excited state with
constant wavefunction and a tower of additional excited states
with non-constant wavefunctions''. However, some other modes are
present which have been discarded in this analysis. When plugging
the relation (41) of \cite{LP} into the equations of motion, the authors
implicitly assume that  their function of proportionality cannot
depend on the mode mass, which is an unnecessary restriction. They
then derive an action (59) for their set of modes, but since their
set of modes is not complete, the resulting action does not
necessarily give the correct inner product for the modes.

Even still, one could still argue that if a ghost were present
among this restricted set of modes, it should still be present
when the full set is properly considered. To see its effect, we
concentrate for now on the ghost found in \cite{LP}. Following the
prescription of \cite{LP}, the ghost mode seems to have its origin
in the two mixing modes $\psi_1$ and $\tilde{\psi}_2$ of mass
$m^2=16 g^2$ which couple in the action as
\ba
    \mathcal{L}_{\psi_1,\tilde{\psi}_2} =&&\mathcal{A}\, \psi_1
    \(\Box-16g^2\)\psi_1
    +\mathcal{C}\, \tilde{\psi}_2 \(\Box-16g^2\)\tilde{\psi}_2\\
    &+&\mathcal{B}\, \tilde{\psi}_2 \(\Box-16g^2\)\psi_1
    +\mathcal{B}\, \psi_1 \(\Box-16g^2\)\tilde{\psi}_2.\nn
\ea
Due to a typo, the suggested way to diagonalize this action
(ansatz (63)) does not get rid of the cross-terms. Instead of
eq.(63), the correct sign for the change of variables should have
been
\ba
    \phi_\pm=\frac{1}{\sqrt{2}}\(\(\mbox{\boldmath$\mp$}d+\sqrt{1+d^2}\)\psi_1\pm
    \tilde{\psi}_2\), \label{change variable}
\ea
with $d=(\mathcal{C}-\mathcal{A})/2B$. This gives rise to an
action where both $\phi_+$ and $\phi_-$ come in with a positive
kinetic term for any values of the parameters. To convince
ourselves, we may perform the equivalent change of variables:
\ba
    \psi_1=\sqrt{\frac{C}{2}}\(\phi_++\phi_-\),\hspace{10pt}\tilde{\psi}_2=
    \sqrt{\frac{A}{2}}\(\phi_+-\phi_-\),
\ea
so that the action simplifies to
\ba
    \mathcal{L}_{\psi_1,\tilde{\psi}_2} =\mathcal{K}_{+}\, \phi_+
    \(\Box-16g^2\)\phi_+ +\mathcal{K}_{-}\, \phi_-
    \(\Box-16g^2\)\phi_-, \ea with \ba \mathcal{K}_\pm&=&\mathcal{A}\,
    \mathcal{C}\pm \sqrt{\mathcal{A}\,
    \mathcal{C}}\, \mathcal{B}\nn\\
    &=&\sqrt{\frac{1}{12}(1+10x^2+x^4)}\(\sqrt{(1+x^2)^2+8x^2}\pm
    (1+x^2)\)>0,\nn
\ea
where we wrote $x=4g/q$ and used their expressions (60)-(62) for
$\mathcal{A}, \mathcal{B}, \mathcal{C}$, so that both
$\mathcal{K}_+$ and $\mathcal{K}_-$ are {\it positive for all
values of the parameter space}. Thus we do not recover the same
result as the one argued in \cite{LP}. The presence of a ghost in
their analysis seems to be due to a sign error in the change of
variable \eqref{change variable}.

In section \ref{sec action}, we re-derive the action exactly for the entire
set of modes in the theory, working in the more physical comoving
gauge, and find the stable action which reproduces the field
equations for the fluctuations.

\section{Decoupled Variables in Different Gauges}
\label{app decoupled variables}

In this appendix we relate the variables $(\tilde{\chi},\tilde
\psi, \tilde f)$ --- in terms of which the fluctuation equations
decouple --- to the fluctuation variables in comoving and
longitudinal gauges.

\subsubsection*{Comoving gauge}

In comoving gauge we have $2A^{(c)} = - \Psi$ and so using the
gauge transformations \eqref{long to com}, we have
\ba
    \tilde f &=& \frac{\kappa q}{\cU}\, e^{\cx} \(
    B^{(c)}+\frac 12 \Phi^{(c)}\)\nn\\
    \tilde \chi &=& -\frac{\sqrt{3}}{4 \cU \cV}\Bigg[ (2 \,\cU^2
    + \cx' \cz' ) \Phi^{(c)}  - 2 \, \cU^2 \Psi^{(c)} + 2 \cx'  \cz'  B^{(c)} \Bigg]\nn \\
    \tilde \psi &=& \left( \frac{2\cx'-3
    \cz'}{4\cV} \right) \,\Phi^{(c)} + \left[ \frac{3\cV}{ (2
    \cy'+\cz')}+\frac{3\cz'}{4\cV}\right] \Psi^{(c)}
     -\left[ \frac{2\cV}{(2 \cy'+
    \cz')}-\frac{ \cx'}{\cV} \right] B^{(c)}  \, ,\nn \\
\ea
where primes denote differentiation with respect to $\eta$.

\subsubsection*{Longitudinal gauge}

In longitudinal gauge our notation is, $\Phi^{(l)}=-f/2$, and
$B^{(l)} = \Psi^{(l)} - \xi^{(l)}/2$. Using the gauge
transformation \eqref{long to com} then leads to
\ba
    \tilde f &=& \frac{\kappa q}{\cU}\, e^{\cx}\left[ \(B^{(l)}+\frac 12
    \Phi^{(l)}\)-\(\frac{\cA'}{\cA}+\frac12
    \varphi_0'\)\frac{\cA_\theta^{(l)}}{a_\theta'}
    \right]\nn\\
    \tilde \chi &=&-\frac{\sqrt{3}}{4 \cU \cV}\Bigg[ (2 \, \cU^2 +\cx' \cz')
    \Phi^{(l)} -2 \, \cU^2 \Psi^{(l)} +  2 \cx'  \cz'  B^{(l)}  \nn \\
    &&-\(2 \, \cU^2 \left[ \varphi_0' + \frac{2\cW'}{\cW} \right] +\cx' \cz' \,
    \left[ \frac{2\cA'}{\cA}+\varphi_0' \right]\)\frac{\cA_\theta^{(l)}}
    {a_\theta'}\Bigg]\nn \\
    \tilde \psi &=& \left( \frac{2\cx'-3\cz'}{4\cV} \right)
    \,\(\Phi^{(l)} -  \varphi_0' \frac{\cA_\theta^{(l)}}{a_\theta'}\)
    +\(\frac{3\cV}{ (2 \cy'+ \cz')}+\frac{3\cz'}{4\cV}\)\(\Psi^{(l)}
    +\frac{2\cW'}{\cW}\frac{\cA_\theta^{(l)}}{a_\theta'}\)\nn\\
    &&-\(\frac{2\cV}{ (2 \cy'+\cz')}-\frac{\cx'}{\cV}
    \)\(B^{(l)}-\frac{\cA'}{\cA}\frac{\cA_\theta^{(l)}}
    {a_\theta'}\), \nn
\ea
where $\cA_\theta^{(l)}$ should be related to the variables
$(B_{(l)},\Phi_{(l)},\Psi_{(l)})$, (or $(\chi_{(l)}, \Gamma_{(l)},
f_{(l)})$) using the perturbed equation \eqref{SolnForA}. We note
that eq.\eqref{SolnForA} gives an expression for
$\cA_\theta^{(l)}$ up to an integration constant $\Omega(\rho)$.
However $\Omega$ only contributes to the massless part of
$\cA_\theta^{(l)}$, for which it can be interpreted as a
redefinition of the background quantities. We can therefore set
$\Omega$ to zero in what follows.

For ease of reference we also give here the specialization of the
above relations for the case of conical singularities, for which
we have:
\ba
    \tilde \chi&=&\frac{\sqrt{3}}{4} \, \chi_{(l)} \nn\\
    \tilde f &=& e^{-\cx} \Bigg[ -\frac{\cU}{\kappa q} \, f^{(l)}
    +\(\frac{3\cU}{4 \kappa q} + \frac{\cx'\cy'}{4 \kappa q \,\cU }
    \)\chi^{(l)}\\
    &&\qquad -\(\frac{\cU}{2 \kappa q} + \frac{\cx'\cy'}{2 \kappa q \cU}
    \) \Gamma^{(l)} - \frac{\cx'}{2\kappa q \cU} \, {\Gamma^{(l)}}'  \Bigg] \nn \\
    \tilde \psi &=&-\frac{\cy'}{4\cU} \, \chi^{(l)} + \frac{1}{\cU\cy'}
    \( -\frac{2\,g^2}{\k} \, e^{2\cy} + \cU^2 \) \Gamma^{(l)}
     +\frac{1}{2\cU} {\Gamma^{(l)}}'\, . \nn
\ea
For consistency, one can check that the field equations (\ref{eq
talpha}, \ref{eq tphi}, \ref{eq tpsi}) in the conical case
together with this change of variable give back the equations in
longitudinal gauge (\ref{chiEqn}, \ref{GammaEqn}, \ref{NewfEqn}).

\section{Some Special Functions}
\label{app special functions}

In this Appendix we record some useful properties of the special
functions which are used in the main text. In our conventions the
Legendre functions are related to Hypergeometric functions by
\bea
    P_\nu(z) &=& F \left[ -\nu, 1 + \nu; 1 ; \frac12(1-z) \right]
    \nn\\
    Q_\nu(z) &=& \frac{\sqrt\pi \, \Gamma(1+\nu)}{(2z)^{1+\nu}
    \Gamma\left( \frac32 + \nu \right)} \, F \left[ 1 +
    \frac{\nu}{2}, \frac12 + \frac{\nu}{2}; \frac32 + \nu; z^{-2}
    \right] \,,\quad \nu \ne -\frac32 , -\frac52, \dots
    \nn
\eea

The asymptotic form of the Legendre functions as $z \to \pm 1$ can
be found from well-established properties of the hypergeometric
functions, most notably its series definition
\be
    F[a,b; c; z] = 1 + \frac{ab}{c} \, z +
    \frac{a(a+1)b(b+1)}{c(c+1)} \, \frac{z^2}{2!} + \cdots \,,
\ee
provided $c \ne 0, -1, -2, \dots$. The potential singularities of
$F[a,b;c;z]$ lie at $z=\pm1$ and $z=\infty$, and our interest
is in particular its behaviour at $z =\pm1$. A useful identity for
identifying these behaviours is
\bea \label{HyperGeoID}
    F[a,b; c;u] &=& \frac{\Gamma(c) \Gamma(c-a-b)}{\Gamma(c-b)
    \Gamma(c-a)} \; F[a,b; 1+a+b-c;1-u] \\
    && \qquad + (1-u)^{c-a-b} \left[
    \frac{\Gamma(c) \Gamma(a+b-c)}{\Gamma(a) \Gamma(b)} \right]
    F[c-a,c-b; c-a-b+1; 1-u] \,. \nn
\eea
This expression shows that $F[a,b;c;z]$ is finite as $z\rightarrow
1$ provided $c>a+b$, it diverges there logarithmically if $c=a+b$,
and it can be more singular if $c < a + b$:
\ba
    && F[a,b;c;1]= \frac{\Gamma(c)\, \Gamma(c-a-b)}{\Gamma(c-a)\,
    \Gamma(c-b)},\ \ \ \hbox{ if }\ \ c>a+b, \ a,b \notin \mathbb{Z}_-
    \nn\\
    && F[a,b;a+b;z]_{\ \ \overrightarrow{z \rightarrow 1}}\ \
    \frac{\Gamma(a+b)}{\Gamma(a)\Gamma(b)} \log (1-z), \ \ \hbox{ if }\  \
    a,b \notin \mathbb{Z}_- \nn\nn\\
    &&  F[-n,b;c;1]=\frac{(c-b)_n}{(c)_n}  \ \ n \in \mathbb{N} \,. \nn
\ea
Here $\Gamma$ is Euler's Gamma function $\Gamma(n)=(n-1)!$,
defined for complex arguments by $\Gamma(x) = \int_0^\infty \d u\
u^{x-1} e^{-u}$.  $(\cdots)_n$ is the Pochhammer symbol, defined
by $(x)_n=\frac{\Gamma(x+n)}{\Gamma(x)}$.

For instance, using eq.~\pref{HyperGeoID} with $a = -\nu$, $b =
1+\nu$, $c = 1$ and $u = \frac12(1-z)$ leads (provided $\nu \ne
0,1,2,\dots$) to the following asymptotic expression
\bea
    P_\nu(z) &=& - \frac{\ln \left[ \frac12(1+z) \right]}{\Gamma(-\nu) \Gamma(1+\nu)}
    \; F \left[ -\nu, 1+ \nu \,; 1 \,; \frac12(1 + z) \right]  +O(1)\nn\\
    &=& - \frac{\ln \left[ \frac12(1+z) \right]}{\Gamma(-\nu) \Gamma(1+\nu)}
    \;
    +O(1)
    \quad \hbox{as $z \to -1$}\,,
\eea
%
If, on the other hand, $\nu = \ell = 0,1,\dots$,
then the hypergeometric series terminates and $P_\ell(z)$ is
bounded at {\it both} $z = \pm 1$. This leads in the usual way to
the Legendre polynomials, for which with $P_0(z) = 1$ and:
\be
    P_\ell(z) = \frac{1}{2^\ell \ell!} \, \left( \frac{\exd}{\exd
    z} \right)^\ell \Bigl( z^2 - 1 \Bigr)^\ell \,, \quad \ell =
    1,2,\dots \,,
\ee
so that $P_1(z) = z$, $P_2(z) = \frac12 (3z^2 - 1)$, $P_3(z) =
\frac12 \,z (5z^2 - 3)$, and so on.

Similarly, using $a = 1 + \frac{\nu}{2}$, $b = \frac12 +
\frac{\nu}{2}$, $c = \frac32 + \nu$ and $u = z^{-2}$ in the
identity \pref{HyperGeoID}, leads to
\bea
    Q_\nu(z) &=& \frac{\sqrt\pi \, \Gamma(1+\nu)}{(2z)^{1+\nu}
    \Gamma\left(1+\frac{\nu}{2} \right) \Gamma\left( \frac12
    + \frac{\nu}{2} \right)} \Bigl[ - \ln \left( 1 - z^{-2} \right)
    \Bigr] \, F \left[ 1 + \frac{\nu}{2}, \frac12 + \frac{\nu}{2} \,;
    1 \,; 1 - z^{-2} \right]  +O(1)\nn \\
%
    &=& \pm\, \frac{1}{2 (\pm1)^{\nu}} \, \Bigl[ - \ln \left( 1 - z^{-2} \right)
    + O \left( 1 \right) \Bigr] \quad
    \hbox{as $z \to \pm 1$} \,,
\eea
where we have used properties of the Gamma function to provide further
simplifications.
This function clearly has logarithmic singularities at both $z =
1$ and $z = -1$. Notice that in the region of interest, $-1 < z <
1$, we have $1 - z^{-2} < 0$, leading to a function which is
complex. Since $\chi$ is given by the real part of $Q_\nu$ we have
the following expressions for the near-brane singularities of
$\chi$:
\bea
%
    \chi(z) &=& - \frac{C_2}{2} \ln \(1-z\)+O(1)\quad
    \hbox{as $z \to 1$} \nn\\
    &=& - \left\{ \frac{C_1}{\Gamma(-\nu) \Gamma(1+\nu)}
    -\frac{C_2}{2(-1)^\nu}
     \right\} \; \ln(1+z) + O(1)
     \quad \hbox{as $z \to -1$} \,.
\eea

The homogeneous part of the equation for $\Gamma$ is also of
Hypergeometric form, leading to the following general homogeneous
solutions:
\be
    \Gamma_h(z) = C_3 F \left[ 1 + \frac{\nu}{2} \,, \frac12 -
    \frac{\nu}{2} \,; \frac12 ; {z}^{2} \right]
    + C_4 \, z F \left[ \frac32 + \frac{\nu}{2} \,,
    1 - \frac{\nu}{2} \,; \frac32 ;{z}^{2} \right]  \,.
\ee
Notice that $F(a,b;b;x) = F(b,a;b;x) = 1 + a x + \frac12 \, a(a+1)
x^2 + \cdots$ is independent of $b$, and so the two Hypergeometric
functions in the above expression become identical to one another
in the special case $\nu = 0$. In fact, for $\nu = 0$ the
Hypergeometric series can be summed explicitly to give
\be
    F[1,b;b;z^2] = \sum_{k = 0}^\infty z^{2k}  = \frac{1}{1-z^2}\, ,
\ee
showing that in this case the two homogeneous solutions are
$(1-z^2)^{-1}$ and $z (1-z^2)^{-1}$.

The asymptotic behaviour of the Hypergeometric functions as $z \to
\pm 1$ is found by specializing eq.~\pref{HyperGeoID} to the cases
($i$) $a = 1 + \frac{\nu}{2}$, $b = \frac12 - \frac{\nu}{2}$, $c =
\frac12$ and $u = z^2$; and ($ii$) $a = \frac32 + \frac{\nu}{2}$,
$b = 1 - \frac{\nu}{2}$, $c = \frac32$ and $u = z^2$. Keeping in
mind that $F[a,b\,;\epsilon \,;x] = 1 + ab\,x/\epsilon + O(x^2)$,
this leads to
\bea
    F\left[ 1 + \frac{\nu}{2} \,, \frac12 - \frac{\nu}{2} \,;
    \frac12\,; z^2 \right] &=&
    \frac{\sqrt\pi}{\Gamma\left(1+\frac{\nu}{2}\right) \Gamma\left(
    \frac12 - \frac{\nu}{2} \right)} \left[ \frac{1}{1-z^2} -
    \frac{\nu}{2} \left( \frac12 + \frac{\nu}{2} \right) \ln (1 - z^2) +
    O(1) \right] \,, \nn\\
    F\left[ 1 - \frac{\nu}{2} \,, \frac32 + \frac{\nu}{2} \,;
    \frac32 \,; z^2 \right] &=&  \frac{\sqrt\pi}{2\,
    \Gamma\left(1-\frac{\nu}{2}\right) \Gamma\left(
    \frac32 + \frac{\nu}{2} \right)} \left[ \frac{1}{1-z^2} -
    \frac{\nu}{2} \left( \frac12 + \frac{\nu}{2} \right) \ln (1 - z^2) +
    O(1) \right] \,. \nn\\
\eea

\end{document}